\documentclass[acmsmall]{acmart}

\setcopyright{rightsretained}
\acmJournal{PACMHCI}
\acmYear{2023} \acmVolume{7} \acmNumber{CSCW2} \acmArticle{270} \acmMonth{10} \acmPrice{}\acmDOI{10.1145/3610061}

\usepackage{rotating}
\usepackage{graphicx}  
\usepackage{url}
\usepackage{booktabs}  
\usepackage{caption}
\usepackage{subcaption}
\usepackage{tabularx}
\usepackage{tablefootnote}
\usepackage{multirow}
\usepackage[normalem]{ulem}

\graphicspath{{figures/}{pictures/}{images/}{./}}

\newcommand{\addd}[1]{\textcolor{black}{\noindent{#1}}}

\newcommand{\add}[1]{\textcolor{black}{\noindent{#1}}}
\newcommand{\remove}[1]{\textcolor{black}{\unskip}}

\begin{document}

\title[Impact of Provenance-Enabled Media]{Examining the Impact of Provenance-Enabled Media on Trust and Accuracy Perceptions}


\author{K. J. Kevin Feng}
\affiliation{%
  \institution{University of Washington}
  \city{Seattle}
  \country{USA}}
\email{kjfeng@uw.edu}

\author{Nick Ritchie}
\affiliation{%
  \institution{BBC UX\&D}
  \city{London}
  \country{United Kingdom}}
\email{nick.ritchie@bbc.co.uk}

\author{Pia Blumenthal}
\affiliation{%
  \institution{Adobe, Inc.}
  \city{New York}
  \country{USA}}
\email{piab@adobe.com}

\author{Andy Parsons}
\affiliation{%
  \institution{Adobe, Inc.}
  \city{New York}
  \country{USA}}
\email{andyp@adobe.com}

\author{Amy X. Zhang}
\affiliation{%
  \institution{University of Washington}
  \city{Seattle}
  \country{USA}}
\email{axz@cs.uw.edu}

\renewcommand{\shortauthors}{K. J. Kevin Feng et al.}

\begin{abstract}

In recent years, industry leaders and researchers have proposed to use technical provenance standards to address visual misinformation spread through digitally altered media. By adding immutable and secure provenance information such as authorship and edit date to media metadata, social media users could potentially better assess the validity of the media they encounter. However, it is unclear how end users would respond to provenance information, or how to best design provenance indicators to be understandable to laypeople. We conducted an online experiment with 595 participants from the US and UK to investigate how provenance information altered users' accuracy perceptions and trust in visual content shared on social media. We found that provenance information often lowered trust and caused users to doubt deceptive media, particularly when it revealed that the media was composited. We additionally tested conditions where the provenance information itself was shown to be incomplete or invalid, and found that these states have a significant impact on participants' accuracy perceptions and trust in media, leading them, in some cases, to disbelieve honest media. Our findings show that provenance, although enlightening, is still not a concept well-understood by users, who confuse \textit{media} credibility with the orthogonal (albeit related) concept of \textit{provenance} credibility. We discuss how design choices may contribute to provenance (mis)understanding, and conclude with implications for usable provenance systems, including clearer interfaces and user education.

\end{abstract}



\maketitle

\section{Introduction}


``Seeing is believing'' is no longer the case on modern social media. From the COVID-19 pandemic \cite{pennycook2020fighting}, to elections around the world \cite{ap-twitter-labels}, to mundane topics such as butter consumption \cite{snopes-butter}, visual misinformation is abound. Manipulated media has consequences that are both substantial and persistent---previous work has shown that forged images can distort memory through false confirmation bias \cite{Strange2011-ye, wade2002picture} and, with the right image, can even threaten the legitimacy of democratic decision-making procedures \cite{resende2019whatsapp}. Even if an individual comes to terms with the truth behind a deceptive edit, the warped perceptions to which they were exposed can linger \cite{sacchi2007changing}. The severity and scale of media manipulation has led many to study interventions and systems to visually flag misleading images and videos to viewers \cite{dias2020emphasizing, jahanbakhsh2021nudges, morris2012tweeting, lazer2018science, ozturk2015combating, zhang2018structured}, powered by algorithmic detection mechanisms \cite{hu2020span, li-fast, cozz-cnn} as well as human crowdsourced ones \cite{birdwatch, bhuiyan2020experts}. However, algorithmic solutions are far from perfect \cite{seo2019trust, brandtzaeg2018journalists}, and the breakneck speed of media distribution often far surpasses the speed at which human fact-checkers can flag content. Further, increasingly sophisticated algorithmic detection techniques come with increasingly sophisticated methods to deceive such techniques \cite{gragnaniello2018analysis, hussain2021adversarial, rozsa2020adversarial}, resulting in an unending cat-and-mouse game. These factors make today's inherently reactive methods insufficient in the fight against large-scale proliferation of fake media. 

Exposing media provenance information to social media users is an appealing proposition in the midst of the visual misinformation crisis. \add{In line with definitions outlined by the Coalition for Content Provenance and Authenticity (C2PA),\footnote{\url{https://c2pa.org/}} we use \textit{provenance} to refer} to basic facts about the origins of a piece of digital media (e.g., image, video), which may include information such as the author and the date, time, and location at which the media was created or edited \add{\cite{c2pa-explainer}}. Technical standards for provenance, such as the C2PA standard \cite{c2pa}, allows media authoring tools to embed cryptographically signed provenance information into media metadata \textit{in tandem with} content creation to create a growing provenance chain as the media is shared and remixed. Information within a chain can then be displayed to users through a user interface (UI) overlaid on social media platforms, providing \textit{on-demand} information to users and fact checkers alike to make a more informed credibility judgement~\cite{emily2022usable}.
While academic scholars and industry practitioners have noted the potential of provenance in improving media trust through transparency and accountability \cite{gundecha2013tool, npp} and have developed technical infrastructures to support provenance \cite{aythora2020multi, arweave, cai,starling, project-origin}, a key question remains: \textit{how does provenance change user perceptions of media in a social media feed?}

In this work, we present an online experiment with 595 US- and UK-based participants to examine the impact of media provenance information on user perception of image and video content in a social media feed. Participants are assigned to a mock Twitter-like feed and rate their perceptions of trust and accuracy of visual content on that feed. They are then exposed to the same feed, but with the addition of media provenance information, and are once again asked to rate perceived trust accuracy. Unlike previous work, our study integrates specifications from the C2PA standard \cite{c2pa-ux} into an interactive feed, which allows us to directly compare participant evaluations of media on a regular feed to one that contains special UIs exposing media provenance information on demand. We refer to the latter as \textit{provenance-enabled} feeds.

First, we examined how introducing provenance information
 changes participants' credibility perceptions of media content. 
\add{Credibility is a multi-faceted construct comprising subjective and objective measurements \cite{bhuiyan2020experts, hilligoss2008developing}. Because its dimensions are too complex to fully capture in one study without inducing significant participant fatigue, we operationalize it in our study using two measurements: trust and perceived accuracy of a claim. We characterize \textbf{Trust} as a self-reported measure of the degree to which a particular piece of media offers reliable information \cite{heuer2018trust, pennycook2019fighting, epstein2020distrusted}, and
perceived accuracy of a claim, or simply \textbf{Perceived Accuracy}, as the extent to which a participant agrees with the claim associated with a particular piece of media \cite{clayton2020real, kirchner2020countering, jahanbakhsh2021nudges, berry2018believability, riggio1987social}. Given these two dimensions, we ask the following question:}

\begin{itemize}
     \item  \textbf{RQ.1}: Is there a significant difference in trust and perceived accuracy before and after interaction with a provenance-enabled feed?
\end{itemize}

\add{In addition to knowing \textit{whether} users' perceptions shifted, we want to identify the \textit{direction} of shift---towards or away from ground truth---to determine the efficacy of provenance in combating visual misinformation. Prior work has found interventions that reduce belief in both true and false information and are thus unhelpful overall  \cite{dias2020emphasizing, lee2023priming}. We measure the correction of one's perceived accuracy of a claim, or simply \textbf{Correction}, as how much closer one's perception of accuracy for a claim moves towards the ground truth of that claim after an intervention~\cite{brashier2021timing, pennycook2018prior}.}  In the spirit of this, we ask:

\begin{itemize}
    \item \textbf{RQ.2}: Does exposure to provenance information\addd{, as delivered through an in-feed UI,} correct misled judgements of truth in deceptive media (both explicitly deceptive through heavy editing or subtlely deceptive via miscaptioning), and preserve correct judgements of truth in honest media?
\end{itemize}


\remove{To measure trust and truth judgements, we collected ratings on 1) how much the participant trusts a particular piece of media to offer reliable information, and 2) how much the participant agrees with a claim (which may be true or false) made about what the media purports to depict. We refer to these two measures as ``trust'' and ``claim agreement'' (we simplify the latter to simply ``agreement''). }

\add{There will be messiness in any real-world deployment of a new technical standard (e.g., HTTPS \cite{bernhard2019https}) as a result of uneven adoption, differences in implementation, and potential malicious and evasive actors \cite{emily2022usable}. Thus, we expect that the provenance chains of a non-trivial quantity of media would be incomplete or invalid in some way \cite{c2pa-ux, c2pa-explainer} for a significant period of time after initial launch. In our provenance-enabled feeds, we vary some of the media items to show incomplete or invalid provenance and measure changes in participant perception. In particular, we ask:}

\begin{itemize}
    \item \textbf{RQ.3}: Are there significant differences in changes in trust, perceived accuracy, and correction upon introduction of incomplete and invalid provenance states?
\end{itemize}

\add{Finally, the concept of a verified provenance chain, as well as the term ``provenance'' itself, may be unfamiliar to many. In social media settings where screen real estate is limited, icons and succinct language are often used to represent standards, such as the AdChoices program \cite{adchoices}. We explore the impact of varying provenance terminology and indicator design:}

\begin{itemize}
    \item \textbf{RQ.4}: Are there significant differences in changes in trust, perceived accuracy, and correction, as well as general comprehension, across provenance indicator design and terminology variations?
\end{itemize}


We found that provenance did indeed have a significant effect on credibility perception. Trust and perceived accuracy post-provenance were mostly lower when participants encountered deceptive media. In particular, perceptions of media that had been composited---combined from two or more media sources for deceptive purposes---became less favourable (e.g., lower trust and perceived accuracy), while the direction of change was mixed for non-composited media. Provenance helped correct truth judgements towards deceptive media, but it also ``overcorrected'' in some cases and shifted perceptions further away from the truth in some non-deceptive media. Provenance state also influenced judgements. Media indicating provenance incompleteness or invalidity had lower trust and perceived accuracy than counterparts that did not show those states. This suggests that users struggle to differentiate the related but orthogonal concepts of \textit{provenance} credibility and \textit{content} credibility. \remove{We quantified the changes with a \textit{difference in means} metric, where we subtract pre-provenance trust and perceived accuracy ratings from post-provenance ones (all ratings were on a 5-point Likert scale). A high-level summary of some of our findings is shown in Fig. \ref{fig:visual-summary}.}

\remove{We did not find evidence of significant differences in any measured variables across provenance indicator design and terminology variations. }\add{We did not find evidence of significant differences as a result of design and terminology variations in provenance indicators. }However, participants expressed desires for further clarification of some terminology and more quick, glanceable ways to determine credibility. The latter reveals a tension between the rapid pace of information consumption on social media and the deliberate, investigative reflections prompted by provenance. We conclude with a discussion of our study's design implications on both a granular level with suggestions for future provenance UI designs, as well high-level considerations and future work for deploying usable provenance in practice.


\section{Related Work}
To situate our study and motivate our methods, we review prior literature in credibility signals, visual fact-checking, and media provenance. Unlike previous work, this paper contributes an empirical study on how users' perceptions of media change upon exposure to provenance information in a social media setting, along with how design choices in provenance-bearing UIs may influence those perceptions.

\subsection{Credibility on Online Platforms}
How do users determine credibility online in the first place? Previous work in online news, social media, and e-commerce agree that credibility is relational \cite{chowdhury2020joint, alsmadi2016interaction, luazuaroiu2020consumers, kim2013effects, brown2007word}---i.e. an individual's judgement is heavily influenced by relationships with other users and online entities---, longitudinal \cite{filieri2018makes, metzger2010social, alrubaian2018credibility}---i.e. built up over time through repeated interactions---, and experiential \cite{morris2012tweeting, metzger2013credibility, metzger2010social, zhang2014examining}---users develop their own evaluation heuristics developed through the aforementioned repeated interactions. Specifically on social media, some of these heuristics include reputation, endorsement, consistency, and self-confirmation, obtained from information such as the post author, content source, and number of comments or reviews \cite{metzger2013credibility}. The magnitude of effects from these heuristics, however, can vary based on evaluation task. Shen et al. \cite{shen2019fake} conducted a study with purely fake images and found that participants' past experiences with photo-editing and social media were significant predictors of image credibility evaluation, while most social and heuristic indicators (e.g. source trustworthiness) had no significant impact. However, even if users wanted to leverage information such as content source to make a credibility judgement, they are often obscured or simply not present in the interface \cite{dejavu, liang2018tracing}.

These challenges have given rise to a variety of mechanisms and designs for communicating media credibility \cite{zhang2018structured}. Some actively dissuade users from interacting with content though making an editorial judgement \cite{ozturk2015combating, yaqub2020credibility, sherman2021designing, twitter-manip-label, im2020signals}. However, this approach is not advisable for platforms that attempt to appear impartial \cite{wsj-impartial}. Thus, significant interest lies in content-neutral techniques that nudge users towards relevant articles, useful resources, fact-checking tools, and more \cite{dias2020emphasizing, jahanbakhsh2021nudges, morris2012tweeting, lazer2018science, fourcorners}. 
In a hybrid approach, some platforms have also attempted to crowdsource editorial judgements without casting any liabilities on themselves \cite{birdwatch, mashable-crowd}. For example, Twitter's Birdwatch program \cite{birdwatch} crowdsources ``notes'' from members to identify problematic tweets, but Twitter explicitly states that the notes ``do not represent Twitter's viewpoint and cannot be edited or modified by [Twitter's] teams.'' 

Regardless of approach, literature has shown that effective credibility indicators should be immediate (e.g. shown at first exposure), in context (e.g. in close visual proximity to the content), and specific (e.g. show exactly what makes the content problematic) \cite{sherman2021designing}, leading to calls for interoperable standards for content credibility \cite{zhang2018structured}. Some of these techniques may seem successful at first glance in reducing the sharing of unfounded rumours \cite{lazer2018science} and false content \cite{jahanbakhsh2021nudges}, as well as improving the accuracy of user interpretations \cite{bode2018see, seo2019trust}. These techniques can also have the reverse effect, however, making viewers more ignorant of factual information \cite{jahanbakhsh2021nudges, dias2020emphasizing, garrett2013corrections}. Other factors, such as display mode, can impact media perception as well---literature in end-to-end encrypted messaging found that icon-based disclosures of encryption actually impacted trust negatively, while text-based ones were effective in gaining trust \cite{stransky2021encryption}. Additionally, Yaqub et al. \cite{yaqub2020credibility} have shown that the success of credibility indicators on social media can vary based on demographics and personal characteristics. Besides relying on UI-based interventions alone, researchers have also underscored the importance of educational programs to increase awareness of and provide tools to combat misinformation online \cite{karduni2019, lindsay2019literacy}.

So far, work on credibility signals has mainly focused on supporting the assessment of truthfulness at a particular point in time. Our work extends current literature into the realm of provenance-enabled media, where key information also lies along a temporal axis. 

\subsection{Visual Fact-Checking Techniques}
Fact-checking for information accuracy has been a common practice in news journalism for decades. However, the proliferation of visual mis/disinformation presents many novel challenges. First, images and videos are powerful in their ability to convince: participants who see fictional news accompanied by even a tangentially related photo are more likely to say they remember the event \cite{Strange2011-ye} and share the content on social media \cite{Fenn2019-px} than those who did not see a visual. Furthermore, sophisticated image and video editing techniques, such as ``deepfaking'' \cite{bode2021deepfaking}, or even lightweight modifications via ``cheapfaking'' \cite{la2022cheapfakes}, make it difficult for the naked eye and even automated algorithms to detect manipulations. This calls for efforts to rethink how traditional text-based fact-checking techniques may be expanded to visual media. 

The computer vision community has dedicated significant attention to the area of ``image forensics''---manipulation detection through hardware and software fingerprinting \cite{fridrich, phan, caldelli} as well as comparisons with similar images \cite{cozz-cnn, moreira-cnn, wu-cnn, li-fast, zhou2018learning}. Modern techniques such as feature learning and convolutional neural networks (CNNs) were used to identify images that underwent copy-move and splicing transformations \cite{cozz-cnn, moreira-cnn, wu-cnn, zhou2018learning}, along with datasets and frameworks in this area \cite{wen-ds, nguyen-ds, hu2020span, He-ds}. More classical techniques have also been able to quickly and effectively identify some manipulation procedures. Li et al. were able to effectively detect copy-move forgeries with hierarchical feature point matching \cite{li-fast}. Other work builds off the observation that even though a manipulation is not always recognizable or visible, the statistical properties of an image or video may change upon being edited via JPEG compression \cite{popescu2004statistical, fu2007generalized, fan2003identification}, Colour Filter Array (CFA) interpolation \cite{popescu2005exposing, goljan2015cfa}, contrast and lighting \cite{stamm2010forensic, johnson2007exposing}, and noise \cite{pan2012exposing}. Additionally, web scraping and reverse image search services such as TinEye \cite{tineye} can help reveal similar or exact images that have previously been posted elsewhere on the internet. These techniques can all be used to supplement, or replace, investigators' manual research to create warning labels for the media \cite{twitter-manip-label}.

\remove{Identifying visual misinformation is no easy task, and researchers have recognized the value of collaboration in this process. Matatov et al. present DejaVu \cite{dejavu} to assist journalists in collaboratively addressing visual misinformation. The system looks for near-identical image matches across the web and extends the matches by crawling and indexing rogue social media sites such as 4chan. Journalists can then collaboratively flag results and highlight flags for other journalists. The Eunoma Project \cite{eunomia} aims to approach fact-checking from a different angle; instead of relying on users to fact-check information on their own, the platform provides information cascade visualizations, context checking features, and provenance identification tools for users to determine whether the information they see can be trusted. Users then indicate their trustworthiness, which then turns into crowdsourced trust indicators for other users on the platform.} 
While significant progress has been made in visual fact-checking, development of checker-evading techniques have progressed right alongside it \cite{gragnaniello2018analysis, hussain2021adversarial, rozsa2020adversarial}. Provenance standards circumvent this by inserting a signed (often cryptographic) signature into media metadata as the media is created or edited. Additionally, offering provenance information may be more comprehensible to users than explaining complex algorithmic or crowdsourced approaches. 


\subsection{Surfacing Provenance in Media}
The benefits of surfacing provenance are several. Exposing a piece of media's prior history of creation, editing, and sharing can empower viewers to \textit{proactively} make a truth judgement for themselves instead of \textit{passively} consuming a misinformation label \add{\cite{c2pa-explainer, emily2022usable}}. Moreover, provenance equips content creators and media forensics experts with more tools to identify fake or illegitimately copied media \cite{dejavu}. Over time, provenance-enabled systems can build more trust in media sources by increasing transparency and accountability through structural mechanisms \cite{gundecha2013tool}. 


One major challenge associated with attaching provenance to media is verifying the credibility of the provenance information itself \cite{c2pa-explainer}. As a result, many researchers have turned to cryptographic methods to establish a single source of truth for provenance. Sidnam-Mauch et al. \cite{emily2022usable} argue that cryptographic provenance systems should provide 4 key assurances: authenticating provenance, verifying content is (un)altered, ensuring users are viewing the same content, and creating unalterable records of changes. Several provenance systems and standards based on cryptographic methods have been developed in the last few years and are summarized in Table \ref{t:prov-sys}. 

\begin{table}
\small
\centering
    \begin{tabular}[ht]{p{4.5cm} p{6.5cm} p{2cm}}
    \toprule
    Project & Description & Year Founded\\
    \midrule 
    Arweave \cite{arweave} & Protocol that allows for decentralized saving of media and  websites to create a permanent record & 2017\\
    \midrule
    Four Corners Project \cite{fourcorners} & Open source tool that lets creators add authorship, media backstory, related imagery, and relevant links in an interactive layer on top of their images. & 2018\\
    \midrule
    Content Authenticity Initiative (CAI) \cite{cai} & Community of organizations developing off-the-shelf, open-source tools to integrate content credentials into media. & 2019 \\
    \midrule
    News Provenance Project (NPP) \cite{npp} & Prototype news feed that explores relationship between blockchain-based provenance and user trust in news images. & 2019 \\
    \midrule
    Project Origin \cite{project-origin, aythora2020multi} & Alliance of organizations to create a process where provenance and technical integrity of content can be confirmed. & 2019 \\
    \midrule
    Coalition for Content Provenance and Authenticity (C2PA) \cite{c2pa} & Open technical standard for media authoring tools to embed cryptographic signatures into media metadata. A merger of CAI and Project Origin. & 2020\\
    \midrule
    Starling Labs \cite{starling} & Academic partnership between USC and Stanford that aims to capture, store and verify digital content through cryptographic methods and decentralized web protocols. Incorporates open source tools from C2PA. & 2021\\
    \bottomrule
    \end{tabular}
    \caption{Table of cryptographic provenance initiatives, partially adapted from \cite{emily2022usable}.}
    \label{t:prov-sys}
\end{table}

Of course, cryptography is not the only path to uncovering provenance. Non-cryptographic methods have also been explored. This includes the web-based W3C PROV data model \cite{w3c-prov}, as well as efforts to identify provenance of social media content using formal methods in network theory \cite{barbier2013provenance}, network information and URLs on Twitter \cite{ranganath2013tool}, and user profile information \cite{gundecha2013tool}. The information security community has also developed a number of technical protocols out of interest for globally consistent logs that records change events emitted by various authorities \cite{melara2015coniks, laurie2014ct, ryan2013email, fahl2014stay, nikitin2017chainiac, yu2015detect}. Laurie \cite{laurie2014ct} introduced Certificate Transparency, a certificate distribution mechanism that records Transport Layer Security (TLS) certificates in a public, append-only log for open monitoring. Systems for verifying activities such as software distribution \cite{nikitin2017chainiac, fahl2014stay}, cryptographic key usage \cite{yu2015detect}, and encrypted email \cite{ryan2013email} have also been created, along with the proposition of general-purpose transparency protocols \cite{chen2020reducing, tyagi2021client}. Although not specifically designed with internet media in mind, such protocols may be adapted to any system that contains regular publishing and updating of uniquely identifiable information units.

Researchers have also recognized importance questions surrounding the usability of provenance systems. Sidnam-Mauch et al. \cite{emily2022usable} state that ``usability challenges pose the greatest barriers to implementing provenance tools,'' and that there may be a mismatch between the ideal degree of transparency to news consumers and news publishers. Prior work has also leveraged data visualization to more effectively communicate provenance information, specifically in scientific workflows \cite{yazici2018data, yazici2021usability}. Sherman et al. \cite{sherman2021designing} found that provenance was a key heuristic used to evaluate the authenticity of text, image, and video media when provided. They went on to conduct a large-scale quantitative evaluation of different indicator designs for video provenance and concluded that future designs should be concise, easy to interpret, passive rather than interstitial, and should be highly specific about details such as source and edits. \remove{In our study, we incorporate}\add{Our study incorporates} these principles into our social media prototype\add{, but differs from the work of Sherman et al. in three key ways. First, we explore provenance indicators for both images and videos, whereas Sherman et al. focused exclusively on videos. Second, our inclusion of the C2PA standard allows users to view a \textit{chain} of provenance information associated with a piece of media, rather than one source as per Sherman et al.'s designs. Finally, Sherman et al.'s survey study captured one variable--interpretability of the indicators---whereas we also are concerned with how the indicators change users' credibility judgements of media}. 

Despite prior technical and empirical work on digital media provenance, it is still unclear how end users' perceptions of media are affected by the introduction of provenance information. This may be due to the current lack of systems that implement provenance in-the-wild. In our work, we integrate provenance into an interactive social media feed and study how media perception changes between a regular feed and our provenance-enabled feed.
\add{\subsection{Measurements of Media Perception Online}}
\label{s:rw-measures}
\add{Given the findings surfaced by prior work, it is important to understand common measurements used by researchers to evaluate user perception of online content. One popular measure is the \textit{perceived accuracy} of a text-based headline or claim \cite{clayton2019partisan, clayton2020real, kirchner2020countering, kuru2017motivated, pennycook2017falls, pennycook2018prior, pennycook2019lazy, pennycook2020implied, pennycook2020fighting, jahanbakhsh2021nudges}. Participants are typically shown a piece of text, sometimes with accompanying information such as an image, and are asked rate how accurate they believe it is on a binary true-false scale \cite{jahanbakhsh2021nudges, pennycook2017falls, pennycook2018prior, pennycook2019lazy, pennycook2020implied}, or a 4- or 5-point Likert scale that accounts for more granularity such as ``somewhat accurate'' and ``very accurate'' \cite{clayton2020real, kirchner2020countering}. From accuracy measurements, Pennycook and Rand derived \textit{discernment}, defined as the difference in perceived accuracy or sharing intentions between true and false headlines \cite{pennycook2019lazy}. They then found that simple nudges to think about accuracy can nearly triple discernment levels in participants' sharing intentions, motivating the use of accuracy reminders to combat misinformation \cite{pennycook2020fighting}.}

\add{Another popular measure is \textit{trust} \cite{fisher2016trouble, heuer2018trust, huang2019trust, sherchan2013trust, pennycook2019fighting, epstein2020distrusted}. Sherchan et al. \cite{sherchan2013trust} conducted a review of literature in social and computer sciences to form a definition for social trust in the context of social networks: trust derived from social capital based on interactions between members of a network. Eslami et al. \cite{eslami2017careful} additionally found that user-detected algorithmic bias on social platforms can lead to a breakdown of user trust. The core concept of trust, however, is still not well-defined. Indeed, Fisher noted that despite evolution of literature in measuring trust in news media over a span of 80+ years, there does not exist an agreed-upon definition or measure of trust \cite{fisher2016trouble}. As a result, many studies measure trust in a similar way as accuracy: participants self-report trust on a Likert scale \cite{heuer2018trust, pennycook2019fighting, epstein2020distrusted}.}

\add{Besides perceived accuracy and trust, measures of \textit{believability} \cite{kim2019combating, moravec2018fake}, \textit{credibility} \cite{moravec2018fake, bhuiyan2020experts, hilligoss2008developing},  and \textit{truthfulness} \cite{saeed2022crowdsourced, soprano2021many} have also appeared in prior work. In Kim et al.'s study of fake news articles \cite{kim2013effects}, believability was a composite measure consisting of three 7-point Likert scale ratings of belief in, trust in, and credibility of an article, respectively. Moravec et al. \cite{moravec2018fake} used the same three 7-point ratings but called their composite measurement credibility rather than believability. We note that credibility has also been used interchangeably with trust \cite{huang2019trust}, but prior work has agreed on the use of credibility as a high-level, multi-dimensional construct consisting of believability, fairness, trust, accuracy, reliability, and ``dozens of other concepts and definitions thereof'' \cite{bhuiyan2020experts, hilligoss2008developing}. Bhuiyan et al. \cite{bhuiyan2020experts} specifically included a blend of subjective (e.g. trust) and objective (e.g. accuracy) measurements in their approach to credibility. We take a similar approach in our work. Overall, we observe significant overlaps in prior work's terminology definitions.}

\add{In this study, we draw from past study designs to measure shifts in credibility upon introducing provenance information in social media feeds. Because the multifaceted dimensions of credibility are too numerous to capture in one study, we both subjective measurements (e.g. perceived accuracy of claim) and objective measurements (e.g. actual accuracy of claim) to compute our variables representing credibility.}\\

\section{Methods}
We explore how perceptions of images and videos on social media change upon the introduction of media provenance information. To do this, we deployed an online experiment with a repeated measures design featuring two social media feed prototypes---a regular, no-provenance one and a provenance-enabled one---with participants in the US and UK. Unlike prior work, which provides empirical and theoretical evidence as to why provenance may be helpful to users \cite{emily2022usable, sherman2021designing, gundecha2013tool, yazici2018data, yazici2021usability}, we integrate provenance into an interactive social media environment and directly observe its effects. 

\begin{figure}[t]
    \centering
    \includegraphics[width=1\textwidth]{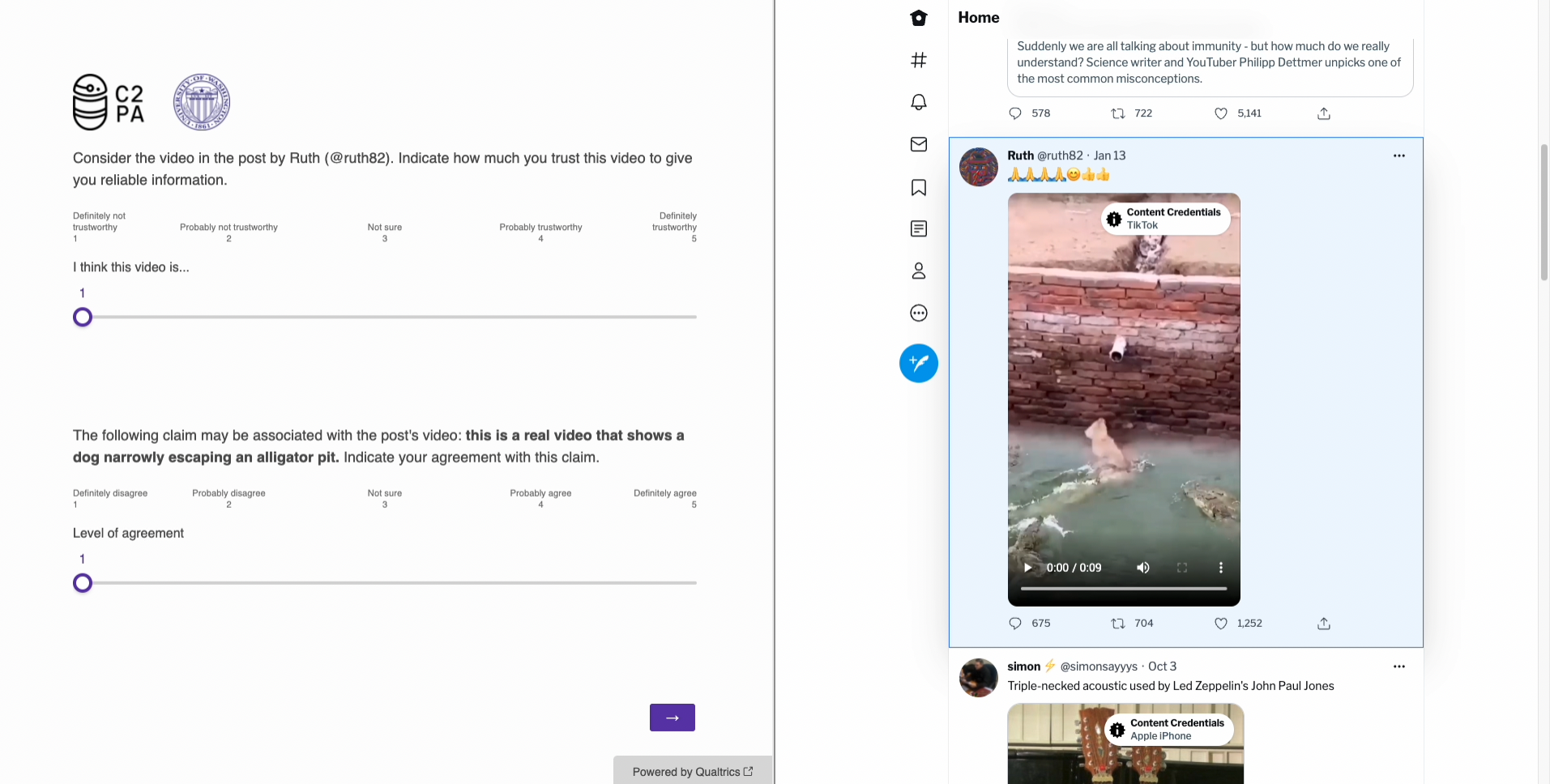}
    \caption{The experiment interface in a side-by-side view during a media evaluation question. The participant responds to the survey on the left while the prototype highlights the relevant post in the feed on the right. Although only one post is highlighted, the participant can still freely interact with the rest of the prototype.}
    \label{fig:system}
\end{figure}

We developed an interface combining a survey and social media prototype for our experiment. Following some preliminary questions about social media use, participants completed two rounds of media evaluations using a mock Twitter-like feed. The first round used a regular feed, while the second used a provenance-enabled feed (see Fig. \ref{fig:l1-l2-b}) with different UI variants (see Fig. \ref{fig:design-spread} in Appendix \ref{a:design-spread}) randomly assigned to participants. The feed automatically scrolled to and highlighted the post associated with the survey question (see Fig. \ref{fig:system}). After both rounds, participants completed an exit survey.

\begin{figure}[t]
    \centering
    \begin{subfigure}[h]{0.3\textheight}
        \centering
        \includegraphics[height=\textwidth]{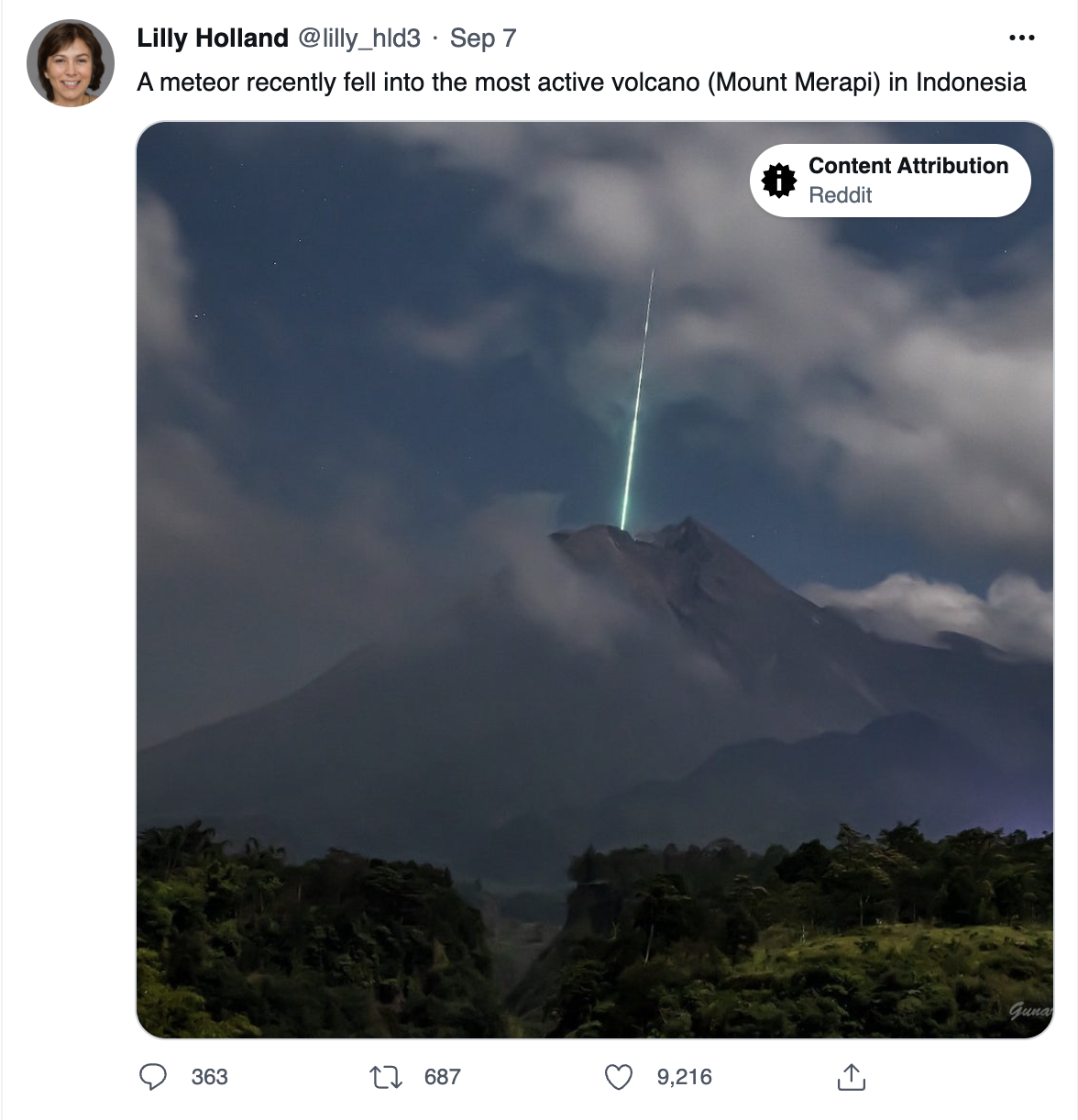}
        \caption{}
        \label{fig:l1-l2-a}
    \end{subfigure}
    \hspace{.1cm}
    \begin{subfigure}[h]{0.3\textheight}
        \centering
        \includegraphics[height=\textwidth]{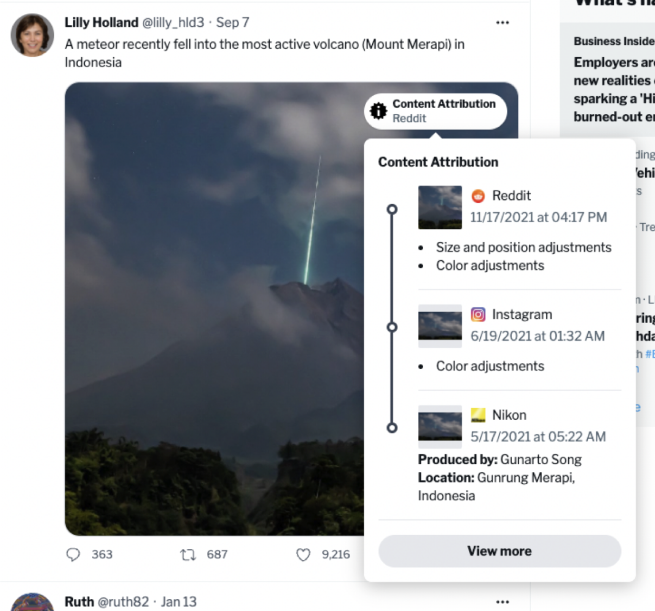}
        \caption{}
        \label{fig:l1-l2-b}
    \end{subfigure}
    \caption{A provenance indicator in a normal state is displayed in \ref{fig:l1-l2-a}. Once the user clicks on it, a provenance details panel will open up, as shown in \ref{fig:l1-l2-b}.}
    \label{fig:l1-l2}
\end{figure}

We further describe our study artifacts and experimental design below.

\subsection{Experiment Interface}
Our experiment interface consisted of a Qualtrics survey, a prototype of a Twitter-like feed, and an overarching system that orchestrates communication between the two.

\subsubsection{Survey}
We collected our data through a Qualtrics survey. The survey began by asking participants about their use of social media platforms, including which platform(s) they use, how frequently they use them, and characteristics that prompt them to question the trustworthiness of a piece of media. 

Afterwards, participants entered two rounds of media evaluation. The first round was the control round of our study, where participants responded to two 5-point Likert scale questions for each post they saw on a no-provenance Twitter feed, similar to a feed they would see on the platform at the time of writing. The two questions measured the two variables (one variable each) used to represent participants' perceptions of media credibility: \textsc{perceived accuracy} and \textsc{trust}. The variables and their corresponding questions are as follows:
\begin{enumerate}
    \item \textsc{perceived accuracy}: \textit{``The following claim may be associated with the post's [image/video]: [claim]. Indicate your agreement with this claim.''}
    \item \textsc{trust}:\textit{``Consider the image in the post by [account name] ([handle]). Indicate how much you trust this image to give you reliable information.''}
\end{enumerate}

Before entering the experimental round, participants were shown a pictorial overview of the provenance UIs we implemented, including their interactive features. In this round, they responded to the same scalar rating questions and saw the same posts as the control round, but did so in a provenance-enabled feed featuring the UIs they were debriefed on earlier. \add{Using the same media across both rounds allowed us to attribute users' changes in perception to the introduction of provenance, eliminating perceptual variations due to differing media as a confounding variable.}

Once they finished, they were taken to an exit survey where they rated their overall understanding of the provenance UIs' functionalities on a 5-point Likert scale, wrote any feedback they had for improving the UIs in a free response question, and completed demographics questions. Select screenshots of the survey can be found in Appendix \ref{a:survey}.

\subsubsection{Social Media Prototype}
\label{s:prototype}
Our social media prototype allowed participants to view and interact with media-embedded posts in a Twitter-like setting. The prototype was a web app built using Svelte\footnote{https://svelte.dev/}. It replicated the appearance of a light-theme Twitter feed and was fully responsive. The number of likes, retweets, and comments were randomized for each post. The order of posts was randomized \textit{between} participants, but were preserved \textit{within} each participant. Besides 3 news posts which we set to be authored by the real Twitter accounts of international news outlets, we generated usernames and handles for the posting accounts using a random name generator, and performed post-hoc editing as needed to ensure there was variance in username styles and genders. We also generated a range of profile photos, selecting those with human faces as well as images and graphics without people. None of the accounts on the feed were verified except for the news outlets. Unlike Twitter, participants could not click into the profile of the posting accounts, nor could they hover over their names and see a profile preview with user statistics such as the number of followers.

Our prototype also integrated analytics tracking. Some of its tracking capabilities included session duration, clicks for various parts of the UI (including open and close clicks where applicable), and interaction timestamps. This allowed us to ensure the participant successfully loaded and interacted with the prototype during the media evaluation rounds.

\begin{figure}[t]
    \centering
    \includegraphics[width=1\textwidth]{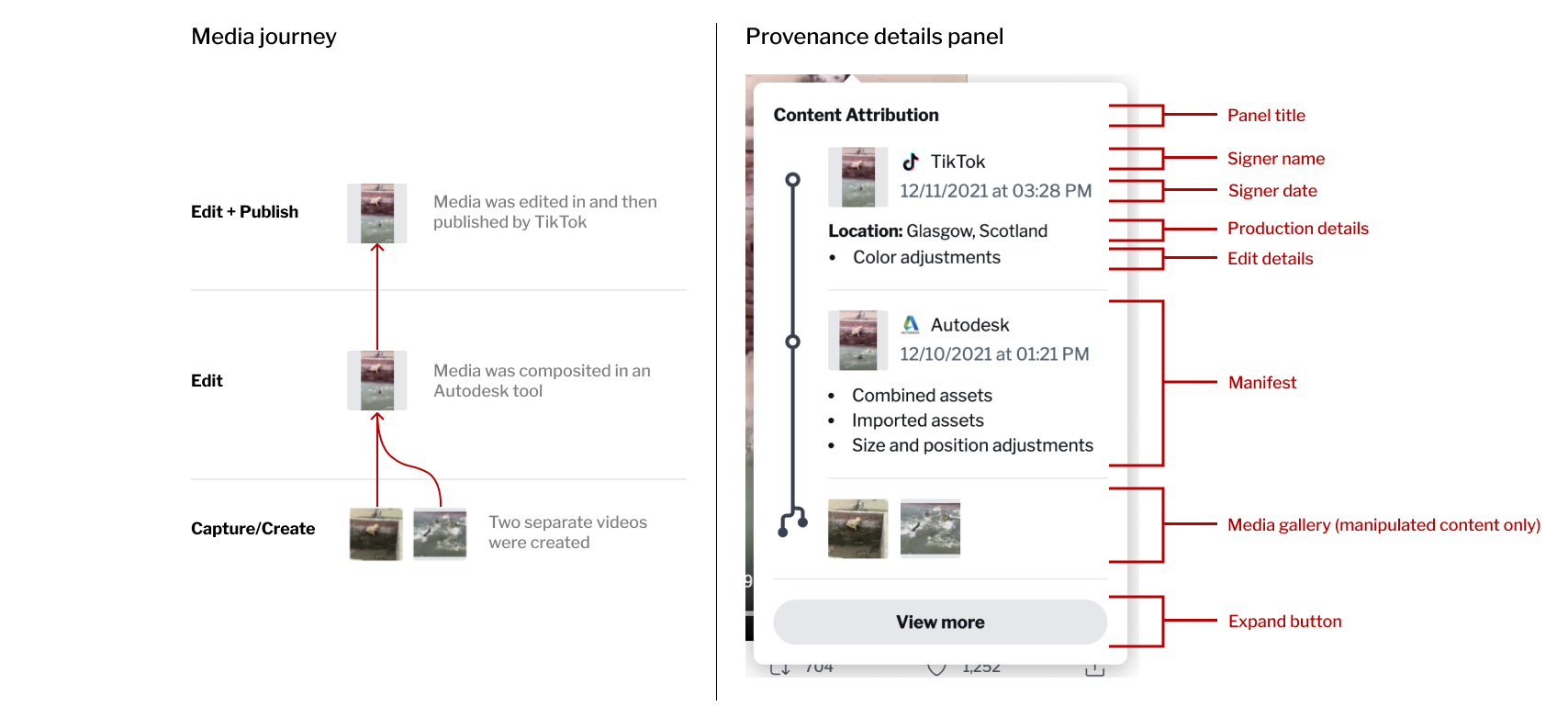}
    \caption{The media journey and the corresponding provenance details panel, including a breakdown of the panel anatomy.}
    \label{fig:l2-anatomy}
\end{figure}

\subsubsection{Provenance UIs}
The feed for the control round did not contain any provenance information, much like a conventional Twitter feed. The provenance-enabled feed in the experimental round, however,  prototypes an implementation of the C2PA standard \cite{c2pa}, an open technical standard that allows media authoring tools to embed cryptographic signatures into the media's metadata. The data can be parsed by the client and displayed to the user through a UI. We implemented the UI in accordance with C2PA's UX guidelines \cite{c2pa-ux}: the user can click on an icon that floats on the media's top right corner---a \textbf{provenance indicator} (see Fig. \ref{fig:l1-l2-a})---to open up a panel displaying a chain of provenance information---a \textbf{provenance details panel} (see Fig. \ref{fig:l1-l2-b}). The anatomy of an example provenance details panel can be seen in Fig. \ref{fig:l2-anatomy}. In our study, we use \textbf{provenance UIs} to refer to the combination of the indicator and details panel. Before entering the experimental round, participants were given a simplified, graphical overview showing that the details panel can be accessed by clicking on the indicator, but were not primed on any specifics within the UIs.

The provenance information can take on three states, as defined by the C2PA standard: \textbf{normal}, \textbf{incomplete}, and \textbf{invalid}. \add{A \textbf{normal} state indicates that there are no abnormalities with the provenance chain. The data carries an \textbf{incomplete} state if the chain was not properly updated after an edit (e.g. an edit was made in a non-C2PA compliant tool). The data is \textbf{invalid} if the chain has been intentionally tampered with, and any portion of the chain before the tampering occurred should be disregarded.} We did not educate participants on the meaning of these states prior to the experimental round to better simulate initial exposure to real-world deployment \addd{through a specific UI representation, to better capture the usability of such a representation}. For example, Twitter did not provide a tutorial for users when introducing new terminology to its warning labels \cite{ap-twitter-labels}, nor did Facebook for its misinformation alerts \cite{fb-alerts}. 

Fig. \ref{fig:l2-anatomy} shows the anatomy of a provenance details panel alongside the corresponding media journey (how the media was created, edited, and published). A details panel consists of a series of manifests, where each manifest consists of a signer name, a signer date, edit details and/or production details. Content that has been composited also has a content gallery showing the original media. According to the C2PA UX guidelines \cite{c2pa-ux}, the Expand button leads to an even more detailed view of the provenance information, but we did not implement that functionality in our prototype. Instead, we tracked its clicks to gauge user interest in further interaction. To craft the manifests within each panel, we used the backstory that Snopes uncovered for each piece of media and fabricated likely signers and details that would fit with the story. For the 3 pieces of news media not from Snopes, we followed the same process but used the image credits in the article instead. We limited the number of manifests in a panel to 2–3, following feedback from our pilot studies that any more would be too confusing.

\subsubsection{Overarching System}
The primary motivation for creating the overarching system was to ensure that participants were actually looking at the posts referred to in the survey as they answered the questions. We implemented the system such that it hosted the survey and prototype simultaneously, and allowed the survey to communicate a question ID to the prototype so the prototype can scroll to and highlight the associated post. 
The system also helped with study mechanics and improving the overall study-taking experience. It randomly assigned participants to one of the many design variants in the experimental round and linked survey response IDs to prototype sessions. 
It full-screened the survey when the prototype was not needed, and put the survey and prototype side-by-side in a 50\% split-screen view when it was. 


\subsection{Media Selection}
\label{s:media-selection}
Rather than fabricating our own images and videos to use in our study, we followed examples from prior works \cite{fan2020juries, pan2022comparing} and used media from real social media posts. Our aim in doing this was to reduce biases we hold as a result of deriving our research questions and hypotheses, which may differ from real-world decisions \cite{kuhberger2002framing} and therefore impose unwarranted effects on the appearance of the media. We drew most of our media from the Snopes fauxtography archives\footnote{https://www.snopes.com/fact-check/category/photos/}, which contains fact-checking investigations of images and videos that have previously been posted on social media. Because the investigations uncover some backstory of the media---for example, what the original image for an edited version looks like---it provided us with reliable sources from which we craft provenance information for that piece of media. Snopes also provided a claim with every entry in the archive (e.g. ``A photograph shows a meteor falling into a volcano''), which we used in our study in one of our measures of media credibility perception. 

We varied the topics of media throughout our feed to minimize potential perceptive bias one may hold towards content of a particular topic; we did this following previous work that suggests the presence of this bias in social media users \cite{flintham2018falling}. We organized our media into three non-overlapping categories of \textsc{media topics} commonly seen in social media feeds: \textbf{news}, \textbf{lifestyle}, and \textbf{national}. \textbf{News} content comprised of mainstream news articles from well-known international news outlets, all published about two months before the deployment of the study. Their header images, titles, and subtitles were displayed in our prototype in the same style as Twitter's Summary Card UI \cite{summary-cards}, while their subtitles were repeated as the posts' caption. \textbf{Lifestyle} content was content intended to trigger intrigue from the viewer, existing primarily for entertainment. The image or video was displayed in the post, along with an accompanying caption that has been posted with the media before elsewhere on the internet. \textbf{National} content was specific to a certain country and included political content as well as content meant to invoke nationalist sentiments. The images and videos, along with the post captions, were displayed in the same way as lifestyle content. The Snopes archive boasted an abundance of lifestyle and national content, but lacked mainstream news articles, so we selected our news media separately. 

Our media also each took on one of two \textsc{edit statuses}: \textbf{non-composited} and \textbf{composited}. \textbf{Non-composited} media did not undergo major graphical modifications---they were either an untouched original copy or had been edited in a way that did not change their visual contents, such as resizing. \textbf{Composited} media had been edited significantly from the original and combined two or more pieces of media together. Additionally, our media were assigned one of two \textsc{claim agreements}, based on their relationship with their Snopes claim: \textbf{agree} and \textbf{disagree}. If a piece of media's claim states what it authentically shows, then the two are considered to agree, and disagree otherwise. For the 3 news articles not from the Snopes archive, we constructed claims for them by objectively stating the content in the articles' header images; thus all news article agreed with their claims. \add{While we discussed \textit{perceived} accuracy in Section \ref{s:rw-measures}, \textsc{claim agreement} here can be thought of as \textit{ground truth} accuracy---the truthful rating one would select if they evaluated the claim while having access to the media's entire backstory. For example, an image that disagrees with its claim is is also inaccurately represented by it\footnote{Consider a composited image depicting a mountain against a blue sky with the claim ``This is an authentic image showing a mountain against a blue sky.'' The image was in fact originally one of a mountain against a gray sky. The claim therefore inaccurately describes the image's authenticity. Someone who has access to this information via provenance should truthfully select 1 (Definitely disagree) while evaluating the claim for the image.}. We use this objective measure to ground our subjective measures, as prior work has done when evaluating content credibility \cite{bhuiyan2020experts}.}

Each participant saw 9 pieces of media. 7 were images and 2 were videos. Every participant saw the same 6 pieces of news and lifestyle media, while US and UK participants saw 3 pieces of American and British national media, respectively. 7 pieces of media were non-composited, while 2 were composited. 5 pieces of media agreed with their claims, while 4 did not. Both pieces of composited media did not agree with their claims. \add{That said, a piece of media does not have to be composited to disagree with its claim\footnote{Consider the following example from our media collection: an authentic photo of a meteor flying over a volcano is taken at an angle such that it looked like the meteor was heading into the volcano itself. The image was reposted with a caption claiming that a meteor has flown into a volcano, but scientific calculations showed that the meteor likely fell about 10km to the north. This information was corroborated by the original photographer, who said he did not hear any crashing noises indicative of the meteor actually striking the volcano \cite{snopes-volcano}.}---we used 2 pieces of such media in our study.} Descriptions of all media, along with their media properties and additional independent variables, are displayed in Appendix \ref{a:all-media} Table \ref{t:media-table}.

\subsection{Experimental Design}

We conducted a repeated measures study using our experiment interface to examine the differences in media credibility perception between no-provenance and provenance-enabled Twitter-like feeds. Since we also wanted to see whether varying the provenance indicator designs had a significant effect on perception, participants were randomly assigned to a design variation in the provenance-enabled feed. We compensated participants at a rate of \$14.49 USD per hour, in line with the state minimum wage. The average completion time for our study was around 15 minutes. Our study design was reviewed and approved by our institution’s Institutional Review Board (IRB) under protocol ID 00014901.

\subsubsection{Data Validation}

We placed two attention check questions in our survey, immediately before participants started each round of media evaluation. The questions referenced prototype content and instructions, serving as an indicator that the prototype successfully loaded for participants in addition to observing attention. We eliminated responses from participants who answered both attention check questions incorrectly. We also eliminated responses from participants who did not leave any traces of activity or open any sessions on our prototype (as per prototype analytics). Finally, since most participants took more than 10 minutes to complete the study, we eliminated responses that used less than 5 minutes. 

\subsubsection{Participants} We recruited participants from the US and UK between the ages of 18 and 65 using Prolific\footnote{https://www.prolific.co}, which offered more rigorous social media pre-screening criteria than Amazon Mechanical Turk. We recruited 5 participants from each region as a pilot to test out the platform and our interface. We modified the wording of our attention check questions upon feedback. For the main study, we recruited a total of 690 participants (360 from the US, 330 from the UK), resulting in a total of 595 valid responses (298 from the US, 297 from the UK). This exceeded our target of 540 valid responses based on power analysis aiming to detect a power of 0.7 with a significance criterion of $\alpha = 0.05$. We recruited in 8 batches for both regions to ameliorate demographic skews: after each batch, we ran a demographics analysis on our participant pool and compared it against the most recent official census data for each country. We did this to figure out which pre-screening criteria we should adjust to better match the census demographics. We used the 2020 Census for the US \cite{us-census} and the 2011 Census for the UK \cite{uk-census}\add{, as they were the most recent census datasets available for the respective countries at the time of our study}. 

All participants used one or more of Facebook, YouTube, Twitter, Instagram, and Reddit at least once a month, with 87\% reporting daily use. 52\% identified as female, 46\% as male, 1\% as non-binary, and the rest did not disclose. 14\% of participants were between 18 and 24, 46\% between 25 and 44, and 40\% between 45 and 65. 86\% reported as White, 6\% as Black, 5\% as Asian, and 3\% as Mixed Race. Racial breakdown by region is available in Tables \ref{t:demo-us} and \ref{t:demo-uk} in Appendix \ref{a:demo-region}, where region-specific demographics may also be found. All participants except for 1 completed at least high school: 36\% reported their highest educational attainment was a high school degree, 16\% had an associate's degree or equivalent vocational training, 33\% had a Bachelor's degree, 13\% had a Master's degree, and 2\% had a Ph.D..

\subsubsection{Manipulation of Independent Variables}

\begin{figure}[ht]
    \centering
    \includegraphics[width=1\textwidth]{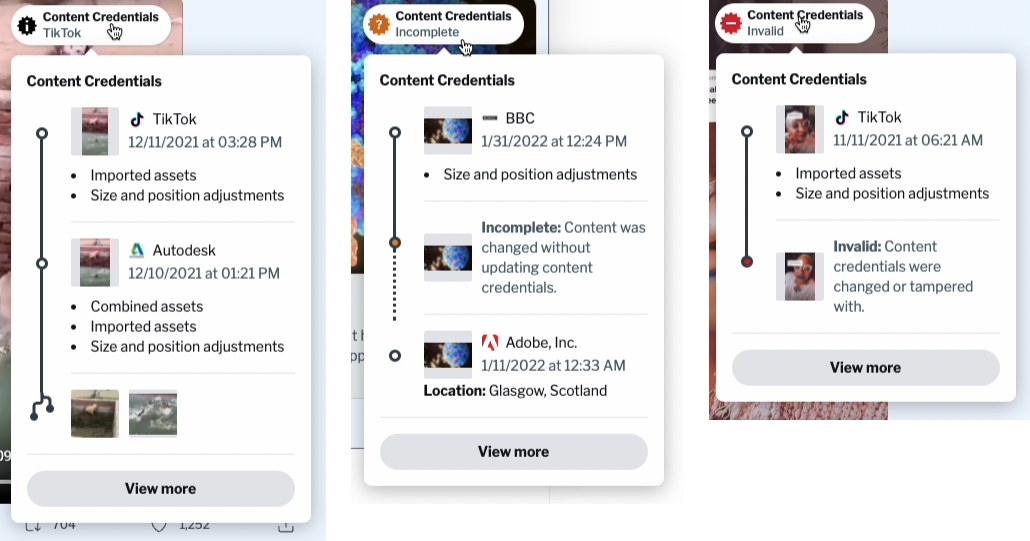}
    \caption{The provenance UIs for various states. From left to right: normal, incomplete, invalid. Note the colour change in the indicator and the addition of incomplete or invalid descriptors in the details panel.}
    \label{fig:stateful-l2}
\end{figure}

As mentioned in Section \ref{s:prototype}, we included 3 \textsc{provenance states} in our study: normal, incomplete, and invalid. State impacted the appearance of the provenance indicator and the contents of the provenance details panel, as shown in Fig. \ref{fig:stateful-l2}. We had specific media we would assign as incomplete or invalid for consistency across feeds. However, since we were interested in isolating the effect of incomplete and invalid states on media credibility perception, not all feeds displayed the full spectrum of states. About half of the feeds saw the presence of all 3 states, while the other half did not have variable states and defaulted to the normal state indicator across all posts. We call the former \textbf{mixed-state feeds} and the latter \textbf{no-state feeds}.

Additionally, we deployed different variations of \textsc{indicator designs} and \textsc{indicator terminology} in our provenance indicators. We used 3 design variations: one was a simple circular icon with an ``i'' in it, one was the icon with a descriptive string (e.g. content provenance), and one was the icon and the descriptive string plus a secondary string. The secondary string bears the name of the most recent signer in the normal state, and the name of the state (e.g. Invalid) otherwise. We used 4 terminology variations in describing provenance in the UIs: ``Content Credentials'', ``Content Attribution'', ``Content History'', and ``Content Provenance.'' These variations appeared in the provenance indicators as well as the panel title in the provenance details panel (see Fig. \ref{fig:l2-anatomy}). In total, we had 18 different variations of indicators in the provenance-enabled feed, shown in Appendix \ref{a:design-spread} Fig. \ref{fig:design-spread}. Our experiment system randomly assigned participants to one of those variations after the control round.  

We also have 3 variables from Section \ref{s:media-selection} that we can treat as independent variables: \textsc{media topic category}, \textsc{edit status}, and \textsc{claim agreement}. All of our independent variables discussed are summarized in Table \ref{t:ivs}. The first 4 variables are considered properties of the media and their assignments to each piece of media in the feeds can be seen in Appendix \ref{a:all-media} Table \ref{t:media-table}.

\begin{table}
\small
\centering
    \begin{tabular}[h]{p{3cm} p{10cm}}
    \toprule
    Independent Variable & Levels\\
    \midrule 
    Media Topic & News, lifestyle, national \\
    Provenance State* & Normal, incomplete, invalid \\
    Edit Status & Non-composited, composited \\
    Claim Agreement & Agree, disagree \\
    Indicator Design & Icon only, icon + string, icon + string + secondary string  \\
    Indicator terminology & Content Credentials, Content Attribution, Content History, Content Provenance \\
    \bottomrule
    \end{tabular}
    \caption{Independent variables and their levels. *Only applies to mixed-state feeds.}
    \label{t:ivs}
\end{table}

\subsubsection{Measurements and Dependent Variables}
We quantitatively measured two ordinal variables that represent \add{credibility perception}: \add{\textsc{perceived accuracy}} and \textsc{trust}. \add{\textsc{Perceived accuracy}} is a participant's response to the question \textit{``The following claim may be associated with the post's [image/video]: [claim]. Indicate your agreement with this claim.''} The claim referred to here is the same claim described in \ref{s:media-selection}---we used the claim provided by Snopes where applicable. The response is in the form of a numerical rating on a 5-point Likert scale, where 1 is \textit{Definitely disagree} and 5 is \textit{Definitely agree}. \add{While we could have asked for accuracy ratings directly, as prior work has done \cite{jahanbakhsh2021nudges, pennycook2019lazy, pennycook2020implied}, we were concerned that doing so would shift the evaluation of accuracy more towards the \textit{factual contents of the claim itself} (which has little relevance to provenance) rather than the \textit{relationship of the claim with the media}. Participants may then be more likely to perform web searches to verify information in the claims, which we wanted to avoid as the claims are published by Snopes and the media-claim relationships are readily available on the internet. Prior work in psychology has used participants' agreement with a piece of information as a reliable proxy for their perceived accuracy \cite{berry2018believability, riggio1987social}. We take inspiration from this and ask for claim agreement instead of accuracy directly to better capture participants' impressions of accuracy via the claims' alignment with their media. }The \textsc{trust} is a participant's response to the question \textit{``Consider the image in the post by [account name] ([handle]). Indicate how much you trust this image to give you reliable information.''} Just like \textsc{perceived accuracy}, the response is a numerical rating on a 5-point Likert scale, where 1 is \textit{Definitely not trustworthy} and 5 is \textit{Definitely trustworthy}. These variables were collected for both the control and experimental rounds. Let us denote \textsc{perceived accuracy} and \textsc{trust} in the control round as $\alpha_c$ and $t_c$, respectively, and in the experiment as $\alpha_e$ and $t_e$.

These measured variables are useful when we check for \textit{whether there exists significant differences} in responses between the two rounds, but not so much in \textit{whether and how those differences change between treatment conditions} should they exist. We process the measured variables into 3 dependent variables we can use for this. 

\begin{enumerate}
    \item \textsc{Change in trust}, denoted by $\Delta t$, is calculated by $t_e - t_c$. 
    \item \textsc{Change in perceived accuracy}, denoted by $\Delta \alpha$, is calculated by $\alpha_e - \alpha_c$.
    \item \textsc{Correction}, denoted as $\alpha_{corr}$, is $\Delta \alpha$ relative to a post's \textit{true agreement/accuracy} ( denoted $A$). Every piece of media has a ground truth \textsc{claim agreement} associated with it (see Section \ref{s:media-selection}). Correction is calculated by $|\alpha_c - A| - |\alpha_e - A|$. Note that we did not set a ground truth for trust due to its subjectivity.
\end{enumerate}

\begin{table}
\small
\centering
    \begin{tabular}[h]{p{5cm} p{4cm} p{3.7cm}}
    \toprule
    Dependent Variable & Formula (If Applicable) & Used In\\
    \midrule 
    Perceived accuracy ($\alpha_c$ and $\alpha_e$) & N/A & All\\
    Trust ($t_c$ and $t_e$) & N/A & All\\
    Change in perceived accuracy ($\Delta \alpha$) & $\alpha_e - \alpha_c$ & RQ.3, RQ.4\\
    Change in trust ($\Delta t$) & $t_e - t_c$ & RQ.3, RQ.4\\
    Correction ($\alpha_{corr}$) & $|\alpha_c - A| - |\alpha_e - A|$ & RQ.2, RQ.3, RQ.4 \\
    Comprehension & N/A & RQ.4 \\
    \bottomrule
    \end{tabular}
    \caption{Dependent variables with their formulas and where it was used. A N/A in the Formula column means the variable was measured directly.}
    \label{t:dvs}
\end{table}

Additionally, it was important for us to know whether the provenance UIs were even comprehensible to participants. We asked the following question in our survey after both media evaluation rounds: \textit{``For the interactive media UIs that were only in second social media feed, did you have a clear idea of what the UIs did? Rate your understanding of their functionality from a scale of Very unclear (1) to Very clear (5).''} RQ.4 probed participant comprehension, which we could not capture well with the other dependent variables, so we captured the Likert scale ratings for the aforementioned question in a variable called \textsc{comprehension} and used it as a dependent variable.

Our dependent variables, along with which research questions they were used for, are summarized in Table \ref{t:dvs}.

\subsubsection{Quantitative Analysis}
\label{s:quant-analysis}
First, we wanted to find out which independent variables were most important in contributing to the participant's rating of \textsc{perceived accuracy} and \textsc{trust}, as well as identify any interaction effects between the variables. To this end, we used the independent variables to create two cumulative link mixed models (CLMMs) with \textsc{perceived accuracy} and \textsc{trust} as the respective response variables, setting \textsc{media topic}, \textsc{provenance state}, \textsc{edit status}, \textsc{claim agreement} as fixed effects, and \textsc{indicator design} and \textsc{indicator terminology} as random effects. We chose to use CLMMs due to the ordinal nature of our measured variables and random effects that may be present in \textsc{Indicator design} and \textsc{Indicator terminology} due to random assignment of participants to indicator variations in the experimental round. 


We then checked for significant differences in \textsc{trust}, \textsc{perceived accuracy}, and \textsc{correction} between the control and experiment rounds to find out if introducing provenance had any significant effects at all. That is, we checked for evidence of non-negligible $\Delta \alpha$, $\Delta t$, and $a_{imp}$. We used a Wilcoxon signed-rank test to do this, as the pre- and post-provenance ratings are always matching samples. If results were significant, we identified the direction and magnitude of change by computing a difference in means across the two groups. We structured our reporting based on the significant interaction effects revealed by our CLMMs: if interactions existed between two variables, we separated the results into the variables' levels to avoid aggregating across those effects. When we check for significant differences in $\Delta \alpha$, $\Delta t$, and $a_{imp}$ \textit{across} independent variable levels, we used the Mann-Whitney U Test for independent groups (computing difference in means where significant) when there were two groups, and the Kruskal-Wallis test (with Dunn's posthoc test and Bonferroni correction for pairwise analysis where significant) when there were more than two. Overall, we chose non-parametric tests for our analyses due to the ordinal nature of our dependent variables.

\subsubsection{Qualitative Analysis}

Our survey included the following free response question as part of the exit survey: \textit{``Do you have any feedback to improve the interactive media UIs that appeared in the second social media feed? Examples include suggestions of a better icon, different name, etc.''} 223 out of the 595 participants gave meaningful responses longer than 5 words. To extract insights from the responses, one author took two passes through the data, first performing an open coding procedure to capture meaningful comments and suggestions, and then organizing the results into broader themes through axial coding. The result was the identification of 4 categories of desiderata for future provenance UIs: \textsc{explainability}, \textsc{interactivity}, \textsc{visibility}, and \textsc{iconography/terminology}.

\section{Results}

Our CLMM models for estimating effects of \textsc{trust} and \textsc{perceived accuracy}, as outlined in Section \ref{s:quant-analysis}, were used to identify significant effects and interaction effects to further explore using hypothesis tests. Both models revealed \textsc{edit status}, \textsc{claim agreement}, \textsc{media topic}, and \textsc{provenance state} as significant main effects. We then examined interactions between the independent variables by crossing them in pairwise combinations and incorporating them into the models. In both models, we found interactions between \textsc{claim agreement} and \textsc{provenance state}, as well as \textsc{edit status} and \textsc{media topic}, to be significant. The coefficients of fixed effects from the two models, along with their standard errors and significance, are shown in Table \ref{t:models}. Note that the independent variables are split up into their levels due to their categorical nature.

\begin{table}[t]
\small
\centering
    \begin{tabular}[h]{p{4cm} p{4cm} p{5cm}}
    \toprule
    Variable Level & Trust Coefficient (Std. Error) & Agreement Coefficient (Std. Error)\\
    \midrule 
    Composited & -0.49 (0.22)* & -0.56 (0.22)*\\
    Disagree & -0.96 (0.10)*** & -0.96 (0.10)***\\
    National & 0.37 (0.10)*** & 0.36 (0.10)***\\
    News & 1.03 (0.09)*** & 0.78 (0.09)***\\
    Invalid & -1.42 (0.15)*** & -1.49 (0.15)***\\
    Normal & 0.45 (0.12)*** & 0.19 (0.15)\\
    Composited * National & -0.91 (0.18)*** & -0.95 (0.18)***\\
    Composited * Normal & -0.26 (0.20) & -0.13 (0.20)\\
    Disagree * National & 0.07 (0.16) & 0.04 (0.16)\\
    Disagree * Invalid & 0.58 (0.19)** & 0.68 (0.19)***\\
    \bottomrule
    \end{tabular}
    \caption{Coefficients, standard errors, and significance from the trust and perceived accuracy CLMMs. $p<0.05$*, $p<0.01$**, $p<0.001$***}
    \label{t:models}
\end{table}

Additionally, past work on credibility indicators and sharing behaviour \cite{yaqub2020credibility} signalled the possibility of significant differences in our responses based on demographic characteristics, which would in turn affect our data aggregation methods during analysis and reporting. We performed hypothesis tests (Mann-Whitney U Test for 2 groups or the Kruskal-Wallis test for 3 or more groups) on data across levels within collected demographic information (gender, age range, country, household income, race, and education attainment). We did not find evidence of significant differences in both trust and perceived accuracy across any demographic levels. 

\subsection{RQ.1: Differences in \textsc{trust} and \textsc{perceived accuracy} in no-provenance and provenance-enabled feeds}
\label{s:rq1}
RQ.1 was concerned with whether provenance information even made a significant difference in participants' ratings of trust and perceived accuracy. To evaluate this, we used the Wilcoxon signed-rank test to compare ratings from the no-provenance and provenance-enabled rounds. To capture the direction and approximate magnitude of change, we computed the difference in means between the two rounds, subtracting the no-provenance rating from the provenance-enabled rating. That is, we computed $\overline{t_e}-\overline{t_c}$ and $\overline{\alpha_e}-\overline{\alpha_c}$, where $\overline{x} = \sum_{i=0}^{n} x_i / n$. Overall, we mostly found significant differences, with trust and perceived accuracy decreasing in most cases. Taking into consideration interaction effects between the variables, we stratified our analysis by \textsc{edit status} and \textsc{media topic}, as well as by \textsc{claim agreement} and \textsc{provenance state}. The results are shown in Tables \ref{t:edit-topic} and \ref{t:agreement-state}, respectively. 


As Table \ref{t:edit-topic} shows, composited content had a larger negative change in trust and perceived accuracy than non-composited content, indicating that participants were informed about the edit through exposure to provenance and changed their judgements accordingly. The differences in $\Delta t$ and $\Delta \alpha$ between non-composited and composited media were statistically significant: ($U = 1741294.0, p<0.001$) for $\Delta t$, ($U = 1620590.5, p<0.001$) for $\Delta \alpha$. We did not observe any cases where the direction of change in trust and perceived accuracy were different. We also did not find evidence for significant differences in trust and perceived accuracy in non-composited national content. 

About half of participants (52\%) saw a no-state feed in which there was no difference in provenance state between feed media and the other 48\% saw mixed-state feeds in which feed media were assigned normal, incomplete, and invalid provenance states. The assignment was fixed, meaning that a piece of media's state in a mixed-state feed remained consistent across all mixed-state feeds. There was evidence for significant differences in all groups, except for no-state media that agreed with their claims. For media with an invalid state and agreed with their claims, trust and perceived accuracy decreased noticeably more sharply than media with the same claim agreement but an incomplete state. This pattern was not observed with content that disagreed with their claims. The only instances of significant positive change in trust and perceived accuracy came from normal-state content that agreed with their claims. Overall, we found significant differences in $\Delta t$ and $\Delta \alpha$ between content that agreed with their claims and ones that did not: ($U = 3109723.5, p<0.001$) for $\Delta t$, ($U = 3012445.0, p<0.001$) for $\Delta \alpha$.

From Table \ref{t:agreement-state}, we noticed differences between ratings of the no-state media (which still bear the normal state provenance indicator by default) and normal state media in the mixed-state feeds. This led us to run a Mann-Whitney U Test to check for significant differences between the normal state media in the mixed-state feeds and those same pieces of media in the no-state feeds. Indeed, rating differences across the two were significant. Compared to the no-state media, the mixed-state media's trust was 0.08 points higher on average ($U = 1056801.5$, $p<0.05$) and its perceived accuracy was 0.09 points higher on average ($U = 1058681.5$, $p<0.05$). This suggests that the visual representation of the normal state may be perceived as more reliable in the presence of incomplete and invalid states compared to simply on its own.

\begin{table}[t]
\small
\centering
    \begin{tabular}[h]{p{1.9cm} p{1.1cm} p{1.1cm} p{1.1cm} p{1.1cm} p{1.1cm} p{1.1cm} p{1.1cm} p{1.1cm}}
    \toprule
    & \multicolumn{4}{c}{Non-composited} & \multicolumn{4}{c}{Composited}\\
    \cmidrule(lr){2-5}
    \cmidrule(lr){6-9}
    Media Topic & $\Delta t$ & $\Delta \alpha$ & $\alpha_{corr}$ & $d_{rem}$ & $\Delta t$ & $\Delta \alpha$ & $\alpha_{corr}$ & $d_{rem}$ \\
    \midrule 
    News & -0.28*** (0.03) & -0.28*** (0.04) & -0.28*** (0.04) & 1.44 (0.03) & N/A & N/A & N/A & N/A\\
    Lifestyle & 0.22*** (0.05) & 0.21*** (0.05) & 0 \hspace{0.2cm} (0.05) & 1.70 (0.04) & -0.75*** (0.08) & -0.92*** (0.08) & 0.92*** (0.08) & 1.17 (0.06) \\
    National & 0.05 (0.05) & 0.01 (0.05) & -0.06 (0.05) & 1.89 (0.04) & -0.65*** (0.06)  & -0.82*** (0.07) & 0.82*** (0.07) & 0.79 (0.05) \\
    \bottomrule
    \end{tabular}
    \caption{Difference in means of \textsc{trust}, \textsc{perceived accuracy}, \textsc{correction}, and remaining distance from pre- to post-provenance, broken down by \textsc{edit status} and \textsc{media topic}. Note that no news content was composited in our study. $p<0.05$*, $p<0.01$**, $p<0.001$***. \add{Standard errors are shown in in parentheses.}}
    \label{t:edit-topic}
\end{table}

\begin{table}[t]
\small
\centering
    \begin{tabular}[h]{p{2.3cm} p{1.1cm} p{1.1cm} p{1.1cm} p{0.9cm} p{1.1cm} p{1.1cm} p{1.1cm} p{0.9cm}}
    \toprule
    & \multicolumn{4}{c}{Agree With Claim} & \multicolumn{4}{c}{Disagree With Claim}\\
    \cmidrule(lr){2-5}
    \cmidrule(lr){6-9}
    Provenance & $\Delta t$ & $\Delta \alpha$ & $\alpha_{corr}$ & $d_{rem}$ & $\Delta t$ & $\Delta \alpha$ & $\alpha_{corr}$ & $d_{rem}$\\
    \midrule 
    No state (no-state feed) & 0.05 (0.04) & 0.02 (0.04) & 0.02 (0.04) & 1.40 (0.03) & -0.24*** (0.05) & -0.35*** (0.05) & 0.30*** (0.05) & 1.23 (0.04) \\
    Normal state & 0.19*** (0.06) & 0.20*** (0.06) & 0.20*** (0.06) & 1.53 (0.04) & -0.15** (0.08) & -0.25*** (0.08) & 0.16* (0.07) & 1.17 (0.06) \\
    Incomplete state & -0.53*** (0.08) & -0.45*** (0.08) & -0.45*** (0.08) & 1.34 (0.06) & -0.66*** (0.09) & -0.86*** (0.11) & 0.86*** (0.11) & 0.76 (0.06) \\
    Invalid state & -1.14*** (0.09) & -1.24*** (0.09) & -1.24*** (0.09) & 2.43 (0.07) & -0.68*** (0.09) & -0.57*** (0.10) & 0.57*** (0.10) & 1.12 (0.07) \\
    \bottomrule
    \end{tabular}
    \caption{Difference in means of \textsc{trust}, \textsc{perceived accuracy}, \textsc{correction}, and remaining distance from pre- to post-provenance, broken down by \textsc{claim agreement} and \textsc{provenance state}. $p<0.05$*, $p<0.01$**, $p<0.001$***. \add{Standard errors are shown in parentheses.}}
    \label{t:agreement-state}
\end{table}

\subsection{RQ.2: Differences in \textsc{correction} upon consumption of provenance information}
\label{s:rq2}
Differences in perceived accuracy ratings pre- and post-provenance show that provenance information has shifted judgement, but for better or for worse? This is what RQ.2 explores. The \textsc{correction} variable, $\alpha_{corr}$, shown in Tables \ref{t:edit-topic} and \ref{t:agreement-state} can offer us some insight by measuring how much closer the participant got to the ground truth rating for each piece of content. We note here that $\alpha_{corr} = \Delta \alpha$ if $\alpha_c - A < 0$ and $\alpha_e - A < 0$. We also include an additional measurement in our tables, remaining distance, that measures how far (in absolute distance) participants' average perceived accuracy ratings (via claim agreement) were from the ground truth agreement rating for each post. We do this to better contextualize $\alpha_{corr}$: even if $\alpha_{corr}$ is positive, its effectiveness may be diminished if the remaining distance is still large. We denote remaining distance with $d_{rem}$.

For non-composited content, the only significant \textsc{correction} identified was in news content. However, this may be partially due to state differences: in mixed-state feeds, each of the 3 pieces of news content took on a different state. The effects of state are further clarified in Table \ref{t:agreement-state}. The significant, positive \textsc{correction} in composited content indicate that provenance was helpful in correcting participants' perceived accuracy by shifting it towards the ground truth rating post-provenance. Lifestyle content had a slightly higher average \textsc{correction} than national content, possibly due to higher skepticism around national content even before provenance was introduced. Indeed, composited lifestyle content had an average pre-provenance perceived accuracy rating of \remove{2.17 }\add{3.09 (std. err = 0.06)} while the same metric for composited national content was \remove{1.79}\add{2.60 (std. err = 0.06)}. 

In no-state feeds, we did not find evidence of significant \textsc{correction} in media that agreed with their claims but did so in content that disagreed. In mixed-state feeds, we found significant and positive \textsc{correction} in all normal-state media, regardless of claim agreement. Incomplete- and invalid-state media had significant and negative \textsc{correction} when media agreed with their claims, but significant and positive \textsc{correction} otherwise. Running a Kruskal-Wallis test and Bonferroni-corrected Dunn posthoc tests revealed that the differences in changes \textit{between} states were significant for all variables discussed. In media that agreed with their claims, incomplete and invalid states actually pushed participants \textit{further away} from the ground truth rating. This is especially noticeable in the invalid state, and is a phenomenon not present in normal-state media nor media that disagreed with their claims. As such, incompleteness and invalidity in provenance may hinder provenance's corrective abilities in truthful content. However, they can aid correction in scenarios with more dubious content. 

To what extent were corrections effective? In general, larger \textsc{correction} corresponded to smaller $d_{rem}$. This was not the case with lifestyle and national content that had been composited: even a larger \textsc{correction} was not enough to close $d_{rem}$ more so for the former over the latter. Additionally, although provenance was sometimes helpful in correcting perceived accuracy, the corrections were relatively slight (usually within one Likert point). Participants also appeared to typically end up around one point away from the ground truth rating, suggesting that they were not hopelessly far off from the truth to begin with.

\subsection{RQ.3: Differences in change in \textsc{trust}, \textsc{perceived accuracy}, and \textsc{correction} upon introduction of incomplete and invalid provenance chains}
To further investigate the effects of the incomplete and invalid states on perception (RQ.3), we measured significant differences in $\Delta t$ and $\Delta \alpha$ between posts in mixed-state feeds and those same posts in no-state feeds. Just like previous analyses, we included $\alpha_{corr}$ as a way to measure whether a participant's change in perceived accuracy was helping them towards or pushing them away from the ground truth rating. This is distinct from our analysis in Sections \ref{s:rq1} and \ref{s:rq2} as we are now measuring differences in responses \textit{across} participants rather than pre- and post-provenance responses \textit{within} participants. We used the Mann-Whitney U Test and its $n$-group generalization, the Kruskal-Wallis test, and found that both incompleteness and invalidity generally lowered trust and perceived accuracy compared to media that did not exhibit those states. We describe observations from each state in more detail.

\subsubsection{Incompleteness}
Out of the two posts on mixed-state feeds that were designated as incomplete, one was a non-composited news article and the other was a composited national image. 

\begin{figure}[t]
    \centering
    \includegraphics[width=0.7\textwidth]{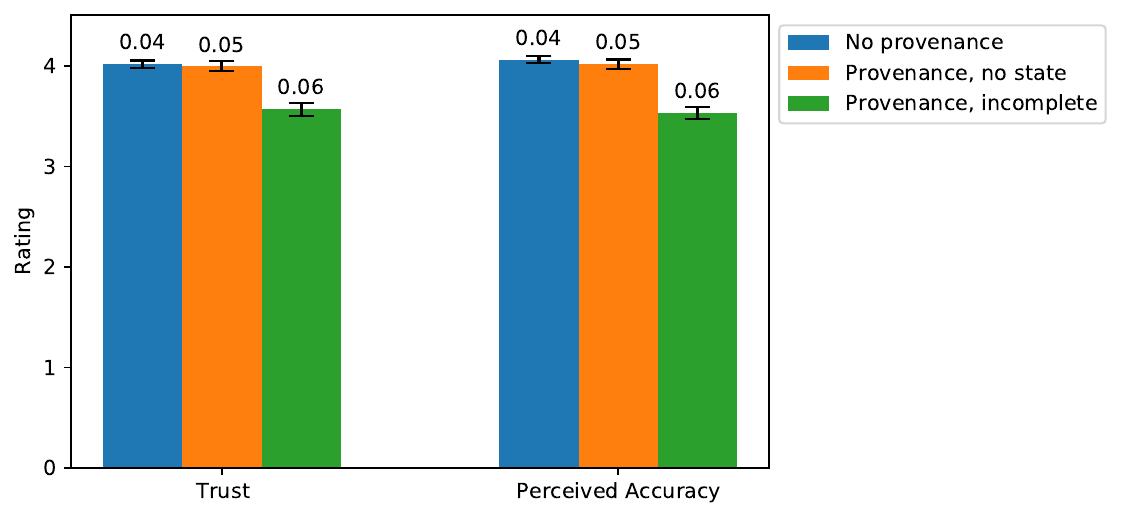}
    \caption{Trust and perceived accuracy ratings for the non-composited news article. Ratings pre-provenance, post-provenance (no state) and post-provenance (incomplete) are shown. Correction, equivalent to the change in perceived accuracy, is not shown explicitly.}
    \label{fig:incomplete-vis}
\end{figure}

For the news article, we observed significant differences in $\Delta t$ ($U = 31637.5, p<0.001$) and $\Delta \alpha$ ($U = 33765.5, p<0.001$) between no-state and incomplete variants of the post. The difference in means from the no-state post to the incomplete post was -0.48 \add{(std. err = 0.07) }and -0.43 \add{(std. err = 0.08) }for $\Delta t$ and $\Delta \alpha = \alpha_{corr}$, respectively. This means that participants who saw the incomplete variant generally trusted the media less and actually moved further away from the truth rating than those who saw the no-state counterpart. Ratings before provenance and after provenance (with and without state) are shown in Fig. \ref{fig:incomplete-vis}.

For the national image, we found no evidence of significant differences in responses between the no-state post and its incomplete variant. 

\subsubsection{Invalidity}
Similar to incompleteness, two posts on mixed-state feeds were designated as invalid. One was a non-composited news article and the other was a non-composited national video.

\begin{figure}[t]
    \centering
    \begin{subfigure}{0.49\textwidth}
        \includegraphics[width=\textwidth]{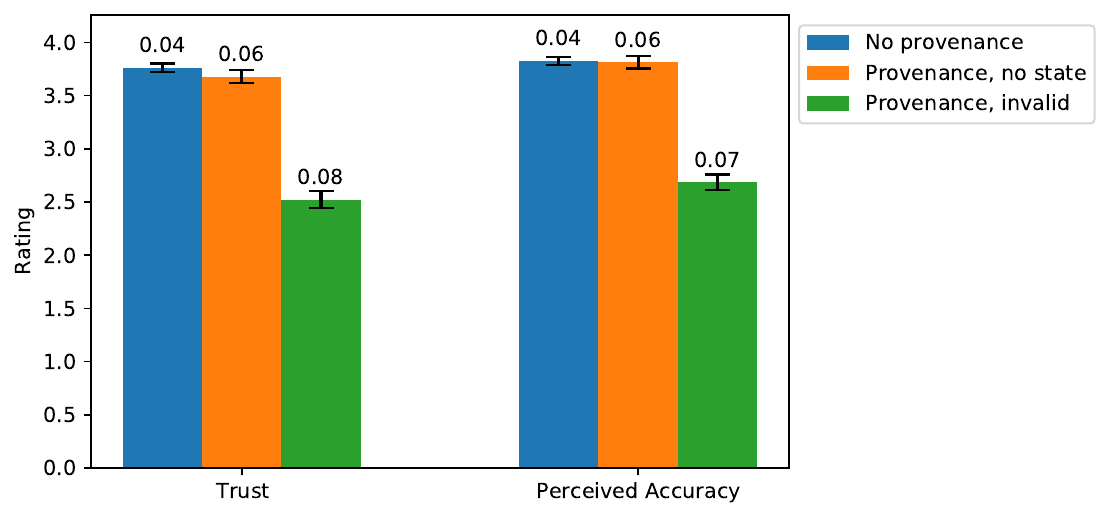}
        \caption{}
        \label{fig:first}
    \end{subfigure}
    \hfill
    \begin{subfigure}{0.49\textwidth}
        \includegraphics[width=\textwidth]{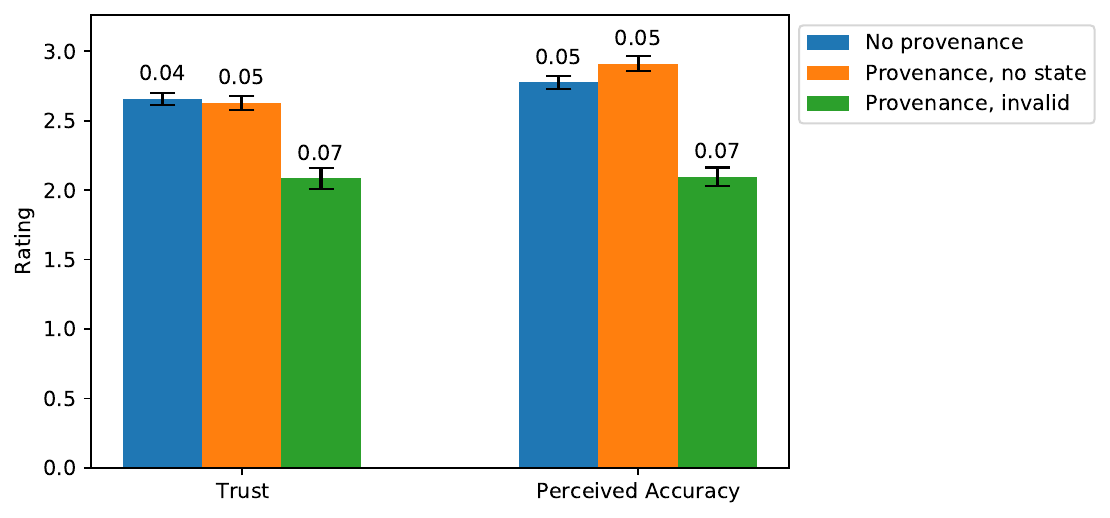}
        \caption{}
        \label{fig:second}
    \end{subfigure}
        
\caption{Trust and perceived accuracy ratings for the non-composited news article (a) and non-composited national video (b). Ratings pre-provenance, post-provenance (no state) and post-provenance (invalid) are shown. Correction, equivalent to the change in perceived accuracy, is not shown explicitly.}
\label{fig:invalid-vis}
\end{figure}

We observed significant differences between no-state and invalid variants of both posts. Here, $\Delta \alpha = \alpha_{corr}$ in both cases. For the news article, the results were ($U = 21544.0, p<0.001$) and ($U = 22288.5, p<0.001$) for $\Delta t$ and $\Delta \alpha = \alpha_{corr}$, respectively. For the video, they were ($U = 25858.0, p<0.001$) and ($U = 31677.5, p<0.001$), respectively. For the news article, the difference in means between the no-state variant and invalid variant were -1.13 \add{(std. err = 0.09) }and -1.16 \add{(std. err = 0.10) }for $\Delta t$ and $\Delta \alpha = \alpha_{corr}$, respectively, a considerably sharper drop than the incomplete state. For the video, they were -0.81 \add{(std. err = 0.08) }and -0.54\add{ (std. err = 0.09)}, respectively. The invalid state amplifies some of the patterns observed in the incomplete state, particularly in the news article: trust and perceived accuracy decreased noticeably in the invalid variant, even when the content aligns with its claim and is non-composited. Ratings before provenance and after provenance (with and without state) for both pieces of media are shown in Fig. \ref{fig:invalid-vis}.

\subsection{RQ.4: Quantitative and qualitative findings across indicator designs and terminology variations}
\subsubsection{Quantitative findings}
RQ.4 explored the possibility of significant differences in dependent variables across indicator designs. Provenance states were represented with some stylistic variations across different indicator designs (shown row-wise in Fig. \ref{fig:design-spread}): some designs simply have a colour-changing icon, while others explicitly show the state name. Using a Kruskal-Wallis test, we found no evidence of significant differences in $\Delta t$ ($H(2) = 0.34, p=0.84$), $\Delta \alpha$ ($H(2) = 0.80, p=0.67$), $\alpha_{corr}$ ($H(2) = 1.98, p=0.37$), and \textsc{comprehension} ($H(2) = 5.91, p=0.05$) across any of the designs. While we did not expect $\Delta t$, $\Delta \alpha$, and $\alpha_{corr}$ to capture the differences in language variations, we failed to find evidence of significant differences in \textsc{comprehension} as well ($H(3) = 1.81, p=0.61$). 

\begin{figure}[t]
    \centering
    \begin{subfigure}{0.49\textwidth}
        \includegraphics[width=\textwidth]{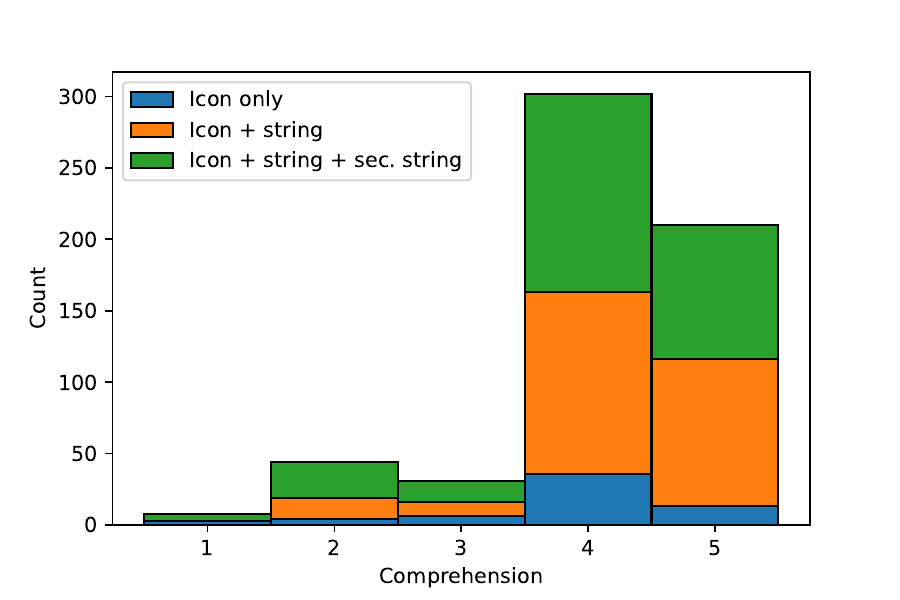}
        \caption{}
        \label{fig:second}
    \end{subfigure}
    \hfill
    \begin{subfigure}{0.49\textwidth}
        \includegraphics[width=\textwidth]{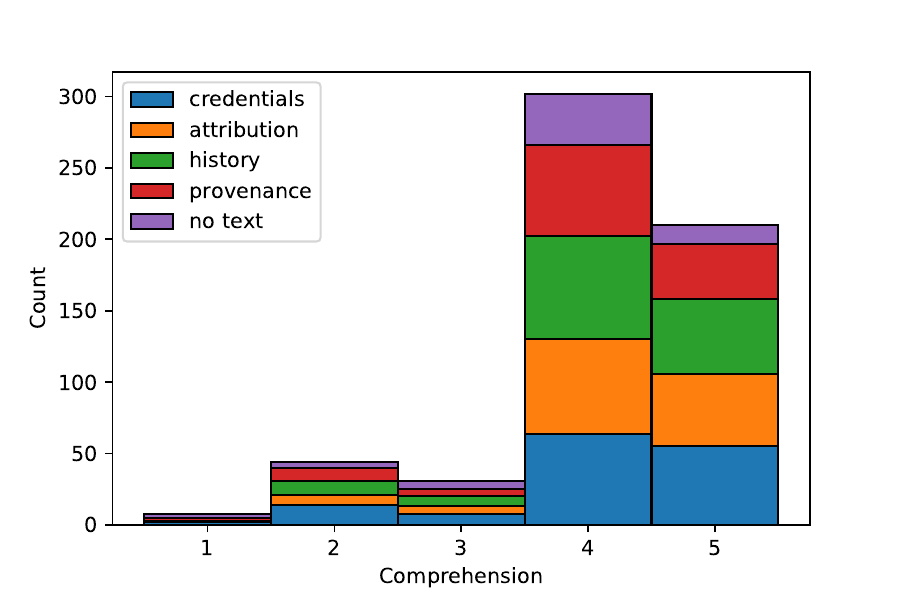}
        \caption{}
        \label{fig:third}
\end{subfigure}
        
\caption{Histograms of participants' comprehension ratings by indicator design (a), and by terminology variation (b). Note that some participants saw only an provenance icon and no text descriptor.}
\label{f:comp}
\end{figure}

Overall, the provenance UIs were rated as well-understood by participants\addd{, although this is merely a \textit{perception} of understanding; we are aware that cognitive biases such as the Dunning Kruger effect may influence ratings}. The mean comprehension rating was 4.11 (std. = 0.90) on a 5-point Likert scale, where 1 was ``Very unclear'' and 5 was ``Very clear''. Fig. \ref{f:comp} shows histograms of the comprehension ratings and its breakdowns by indicator design and terminology variation.

\subsubsection{Qualitative findings}
\label{s:qual}

\begin{table}[t]
\small
\renewcommand{\arraystretch}{1.5}
\centering
    \begin{tabular}[h]{p{0.5cm} | p{5cm} p{7cm}}
    \hline
     & Desiderata & Example Quote\\
    \hline
    \multirow{3}{=}[-25pt]{\begin{sideways}Explainability\end{sideways}} & Explain terminology in UI & ``I wasn’t sure if invalid meant it failed a check or the check produced an invalid result.'' (P152)\\
     & More assertive explanations and terms & ``Change `incomplete credentials' to something more condemning of the information, so `unreliable source' or something similar.'' (P253)\\
     & More explanation of signer-edit relationship & ``I think I understood the `Adobe' information to mean that the image was manipulated but that was not totally clear.'' (P159)\\
     \hline
    \multirow{2}{=}[-11pt]{\begin{sideways}Interactivity\end{sideways}} & Reveal explanations on-demand through interaction & ``I think a tooltip explaining what the tampering means and what all of the other messages mean when you look at the history of the image would be nice'' (P6)\\
     & Enable a separate view with more details & ``A zoom feature that pops out an image in another window / fills screen.'' (P423)\\
    \hline
    \multirow{2}{=}[-25pt]{\begin{sideways}Visibility\end{sideways}} & Make indicator more noticeable & ``I think the icons could be a little bit larger to make them stand out slightly more.'' (P475)\\
     & Avoid blocking content with indicator & ``Less blocking of content would be nice. Maybe more in the corner or more transparent.'' (P9)\\
     & Use colour to attract attention to suspicious content & ``Better icons would be great; green for trustworthy, red for untrustworthy, grey for unsure.'' (P311)\\
     \hline
     \multirow{2}{=}[-22pt]{\begin{sideways}Icons/Terms\end{sideways}} & More specific icon on indicator & ``I would suggest using a magnifying glass icon because this tool allows us to examine the media in more detail.'' (P491)\\
     & Indicator icon should change according to edits & ``Maybe different labels and maybe more clearer icons, especially for importing and major image editing'' (P96)\\
     & Avoid or clearly define jargon & ``Maybe a definition of provenance because I didn't think about what it meant on the first one.'' (P14)\\
     \hline
    \end{tabular}
    \caption{Summary of themes, desiderata, and example quotes from participants' free responses.}
    \label{t:quotes}
\end{table}
Although we did not observe evidence for significant differences across indicator design and terminology variations, it does not imply total impartiality in those areas. Qualitative feedback revealed preferences and design shortcomings not captured quantitatively using our dependent variables. We received qualitative feedback through our free response question, and analyzing it provided takeaways and desiderata for improving the usability of provenance UIs. 223 out of the 595 participants provided meaningful feedback of more than 5 words. We identified four themes in their responses: \textsc{explainability}, \textsc{interactivity}, \textsc{visibility}, and \textsc{iconography/terminology}. A summary of the insights can be found in Table \ref{t:quotes}.

\textbf{\textsc{Explainability}}: participants generally wanted more explanations in the UIs to clarify terminology around both state (e.g. ``invalid'', ``incomplete'') as well as descriptions about the nature of the edit (e.g. ``size and position adjustments''). P136 echoed P152's sentiment in Table \ref{t:quotes}: \textit{``I wasn't sure of the validity when some items were marked "Incomplete." Defining that would help me better determine the validity of the source.''} Other participants suggested \textit{``Something such as `Likely invalid because of \_\_\_\_, or \_\_\_\_ indicates this as likely to be true.'''} and to \textit{``Change `incomplete credentials' to something more condemning of the information, so `unreliable source' or something similar''}, which is shows a slight misinterpretation of state, as potentially unreliable provenance information may not necessarily correspond to an unreliable source. More explanations around how state was determined may help dissipate confusion and reinforce state as an indicator of \textit{provenance} credibility over \textit{media content} credibility. 

Regarding text-based descriptions about the edits, participants wanted more clarity on whether they should still trust content with those edits: \textit{``Better descriptions, instead of "combined media" perhaps "uses elements from several sources" to be more clear that the content is manipulated and not single source.''} Multiple other participants mentioned that they wanted to see more detailed descriptions about the edits, as it was not always clear what the change was. Additionally, participants wanted a stronger link between the signer shown in the provenance details panel affects reliability of the piece of media. P37 stated \textit{``I just wasn't sure what some of the sources indicated. E.g. if Getty Images meant it was a stock photo. Or what it meant for a picture to come from Adobe but without obvious edits.''} This implies that users may automatically associate some signers with particular types of edits, which may or may not be justified. This association is made even more clear by P159: \textit{``I think I understood the `Adobe' information to mean that the image was manipulated but that was not totally clear.''} Some explanations to educate users on the role of the signer may help users avoid unsupported signer assumptions. This is especially important when a signer's reputation may be used a signal of reliability: \textit{``showing that pictures were created by reputable news organizations greatly helped me determine if they were real''} (P216).

\textbf{\textsc{Interactivity}}: to address some of the explainability issues, participants have recommended UI interactions that reveal supplementary information on-demand. A few participants suggested tooltips that provides definitions of terms when one hovers over them\add{: ``I think a tooltip explaining what the tampering means and what all of the other messages mean [...] would be nice'' (P6). Similarly, P327 desired ``clickable explanation[s] of the definition of terms, e.g. `incomplete.'''} Many indicated interest in accessing more provenance information besides those displayed in the details panel and clicked on the ``View More'' button, but the button was not a functional one in this study. Many also had ideas for ways to further explore the media in intermediate steps of the provenance chain: the most commonly-mentioned features were zooming further into the content to inspect it, and an expanded view that open the details panel up into a full screen and allow direct comparisons with other versions. \add{For instance, P423 suggested ``a zoom feature that pops out an image in another
window/fills screen,'' while P328 added that such UIs should contain ``clickable links to the original documents/sources.''} 

Conveniently, the expanded view participants described resembles C2PA's provenance UI \cite{c2pa-ux} at one level of detail higher than the provenance details panel that appeared in our feeds under the indicator. This view was not implemented in our study but can be accessed from the ``View More'' button in deployment-ready versions of the UI. While we did not study this expanded view, we can confirm many design choices and features with our qualitative data. 

\textbf{\textsc{Visibility}}: Some participants mentioned that the provenance indicator was not noticeable enough. A couple mentioned they would not have noticed the indicator if it wasn't for the pictorial overview of the UIs before the second round of media evaluations. P475 suggested larger icons: \textit{``I think the icons could be a little bit larger to make them stand out slightly more,''} which was also echoed by several other participants. Since the indicator was overlaid on top of the media, poor background contrast also presented visibility concerns: \textit{``I guess it could use an icon that contrasts more with the background. I didn't notice it immediately''} (P543). 

Despite some indicator designs having a smaller surface area and potentially having less visibility than others (e.g. the singular icons in Fig. \ref{fig:design-spread}, there was no clear correlation between the design assigned to the participant and them mentioning poor visibility. On the other hand, some participants also thought the indicators were too obstructive. Participants thought the indicator should be \textit{``maybe placed differently/not covering image''} and also suggested increasing opacity to interfere less with the underlying content. This reveals a fine (and potentially subjective) line between acceptable visibility and acceptable obtrusiveness. 

In addition to indicator visibility, many participants also brought up the use of colour and how it can help increase the visibility of edits. P181 wanted edits to be \textit{``color [assigned] to indicate more obviously if things seem suspicious.''} Icon visibility can also be aided by using brighter colour to \textit{``attract attention''} and \textit{``make it stand out more''}. In general, it seems like participants wanted a colour to provide a quick and easy way to tell whether they should trust a piece of media. P340 suggested to use \textit{``a more simple colour coded green, amber, red system of authenticity.''} P311 expressed similar thoughts in Table \ref{t:quotes}. P225 wanted the signal to be even more explicit: \textit{``Manipulated in red letters would be very helpful.''} While clarity is important, we are concerned that using colour this way could misrepresent the relationship between provenance and content. The confusion can be seen in P97's response: \textit{``I like the color change that indicated there could be a problem with the image.''} Here, colour change was indicating a potential issue with  provenance data, not necessarily in the image itself. More user education may be needed before using colour in the way participants suggested.

\textbf{\textsc{Iconography/Terminology}}: Participants noticed the the icon in the provenance indicators were somewhat generic (P93 called it \textit{``just a random one it seems''}) and had suggestions for alternative designs. P301 suggested a question mark, but did not experience the mixed-state feed and was not aware that the question mark was already used in the incomplete state. P491 suggested a magnifying glass, but also acknowledges that \textit{``this may confuse people into thinking it would enlarge the image.''} Dynamic icons were also proposed by P96 in Table \ref{t:quotes}. P267 was more imaginative: \textit{``the symbol could maybe be a small plant growing?''} At least a couple participants initially confused the icon with the three dots ``...'' on the side of the post, which was a non-functional button in our study. Thus, we can see how using recognizable UI visual language in the indicator may be a natural choice but presents challenges when the language overlaps with UI on current platforms. Participants also had suggestions for terminology used on the indicators, particularly those who saw the term ``provenance.'' P184 made a point about how \textit{``Some people might not be familiar with the word `provenance.' `Origin' might be more familiar,''} while P14 wanted \textit{``a definition of provenance because I didn't think about what it meant on the first one.''} This feedback calls for terminology that is simple yet precise, in line with previous findings on provenance indicator designs for videos \cite{sherman2021designing}.

\section{Discussion}
Our results showed that participant ratings of media trust and perceived accuracy before and after exposure to provenance information were different to a statistically significant degree. Additionally, those differences varied with respect to the provenance state and media characteristics. We will now discuss higher-level observations from our findings and pose implications for the design of future provenance-bearing UIs on social media.

\subsection{Permeable boundary between provenance and content credibility}
Ideally, having access to provenance information should increase trust, perceived accuracy, and correction in authentic media (i.e. non-composited content that agrees with its claim) \cite{gundecha2013tool}. For inauthentic media, trust and perceived accuracy should decrease (as long as the claim states what is supposed to be represented), while correction should remain positive \cite{gundecha2013tool, emily2022usable, sherman2021designing}. We see from Table \ref{t:edit-topic} that non-composited content had positive $\Delta t$ and $\Delta \alpha$ in some cases, while the only significant $a_{imp}$ was negative. While the effect of provenance here does not match with the ideal scenario, it is important to note that a couple pieces of non-composited media did not agree with their claims, which lends some justification for the negative changes.


When we introduce provenance state in our analysis, we \remove{can add the additional dimension to our expectations for the ideal scenario. Content that does not display state may see a positive or negative change in trust and perceived accuracy ratings depending on the edits and claims, but correction should increase given the additional insights into the media's backstory. We }expect that participants will be more skeptical of media with provenance incompleteness and invalidity, drawing from previous work on misinformation warnings \cite{ozturk2015combating, seo2019trust}. For incomplete states, there is no additional information given to evaluate perceived accuracy, so we do not expect a change there; however, invalid states (which occur when the provenance chain is tampered with \cite{c2pa-ux}) may justify lowering perceived accuracy due to suspected malicious intent. 
However, we see from Table \ref{t:agreement-state} that correction and change in perceived accuracy are negative for both incomplete and invalid states. That is, provenance had the opposite effect for correction as originally intended---instead of correcting perceptions, it caused participants to stray further from the truth rating. We see a similar pattern with previous work in credibility indicators, where users exposed to the indicators mistakenly saw true content as false \cite{ dias2020emphasizing, garrett2013corrections}. This is particularly noticeable in the invalid state, where an authentic news image garnered significantly less trust and perceived accuracy than its normal state counterparts. In general, incomplete and invalid states both appear to decrease trust and perceived accuracy in content regardless of content claim agreement. 

One likely explanation for this is that participants confuse \textit{content} credibility with \textit{provenance} credibility. Indeed, we see evidence of this confusion in our qualitative analysis. For example, P97, P311, and P340 all provided feedback on the icon colour while implicitly or explicitly referring to the contents of the media rather than the provenance information displayed in the UI. Colour was used to indicate the credibility of the provenance chain but participants appeared to treat it as a glanceable indicator of content credibility. This is not surprising given that the current landscape of credibility indicators mostly involve action-oriented warnings about the content itself \cite{sherman2021designing, twitter-manip-label, jahanbakhsh2021nudges} and the hasty consumption of information on social media \cite{flintham2018falling}. However, content and provenance credibilities are related but orthogonal concepts. \add{For example, a deceptively composited photo (with low content credibility) may bear a normal provenance state (with high provenance credibility) because its provenance information was updated properly at every edit stage. A viewer who is not yet aware of the composition may incorrectly assume that the photo's content is more credible than another that is unedited but has an invalid provenance state.} 

As our example hints, the lack of distinction between content and provenance credibility can lead users to over-trust inauthentic media with a good provenance record or under-trust authentic media with imperfect provenance. Invalid provenance can indicate deceptive intent, but incomplete provenance can arise in benign situations. Malicious actors may intentionally introduce incompleteness to lower perceived credibility or flood a social media feed with red invalid provenance indicators. It is therefore essential to supplement, or even preface, the deployment of provenance-enabled systems with sufficient user education and on differences between content and provenance credibilities.

\subsection{Not all edits are created equal}
\label{s:edits-unequal}
We see from Table \ref{t:edit-topic} that trust and perceived accuracy both decreased while correction was positive in composited content. While those movements are favourable in the context of this study as both pieces composited media were deceptive (i.e. did not agree with their claims), they may not be in all composited media. Compositions can take on a wide variety of forms. Placing a watermark over or a logo in the corner of an image or video is a common act of composition, and while the media has been edited, it can still authentically depict original content. We cannot extrapolate the movements in trust, perceived accuracy, and correction to media in our study to innocent compositions, but we are aware of this possibility given participants' skepticism over content manipulations in our free response question. Universal decrease in trust and perceived accuracy in composited media may present significant challenges in provenance adoption. Content creators who add a logo or watermark to their media may hesitate publishing to provenance-enabled platforms when users will unjustifiably view their content as less reliable and trustworthy. More consideration of how to differentiate routine compositions from malicious ones is needed.

This need is not solely constrained to compositions, but applies to edits more generally. Edits can be graphically intensive but not significantly change the meaning of media. For example, the Pulitzer-winning photo taken by John Filo during the Kent State Massacre in 1970 was anonymously edited for aesthetic reasons before appearing in several high-profile magazines \cite{petapixel}. On the other hand, a small edit gives opportunity to craft an entirely different story about a piece of media. For example, After the French national football team won the World Cup in 2018, a photo circulated on Facebook that appeared to show Black players in the second row at a photo op at the French presidential palace \cite{observers}, leading to accusations of racial discrimination. The photo was later revealed to have been cropped and the original photo showed Black players in the front row. These two examples illustrate how the graphical complexity of an edit may not correspond to its importance in changing the meaning of media, and edits of similar complexities may have different levels of importance. This separation of importance is currently not represented in our provenance UIs. \add{Relatedly, some manipulations may make deceptive intentions clear to users if they appeared in the provenance chain, while others may not and can even further confuse the user. We can envision the future development of a taxonomy of manipulations based on their efficacy in revealing deceptions via provenance.}

\subsection{Design implications}
\label{s:design-implications}
In this study, we placed provenance indicators in the top right corner of media in a post, in line with the C2PA UX specifications \cite{c2pa-ux}. However, there are many alternatives for placement and design of provenance indicators, such as appearing below or in the corner of the media instead of on top of it. This may require changing the indicator design and making it text-based and native to the platform UI. Text-based indicators that conform to platform design guidelines may be perceived as more trustworthy than graphical icons, as demonstrated in previous work on end-to-end encrypted messaging. \cite{stransky2021encryption}. The use of colour is also a factor to consider in indicator design. Currently, many participants have requested bolder use of colour to create more obvious signals of problematic content. However, because content and provenance credibilities are still not well-distinguished concepts, we think using colour in the way participants suggested can actually be misleading. Early deployments of provenance indicators can consider removing colour altogether to sidestep this confusion until the users establish a clearer mental model of provenance and its relationship to content credibility. 

This study's provenance details panels were relatively simple, with provenance chains of no longer than 3 manifests. However, they may be much more complex in practice. As suggested in Section \ref{s:edits-unequal}, highlighting the importance of an edit with respect to visual semantics rather than graphical complexity can help ``cut through the noise.'' The current design of the provenance does not establish visual hierarchy among edits details (see Fig. \ref{fig:stateful-l2}). Designing a hierarchical system for visually displaying edits according to their semantic relevance may be essential for managing complexity in real-world provenance detail panels. 

On a higher level, we observe design challenges stemming from a foundational tension that exists between provenance and current social media norms. Participants' desire for quick and easy ways to tell whether a piece of media is authentic (as they explained in Section \ref{s:qual}), along with current treatment of misinformation labels as warnings, reveal that users are increasingly accustomed to \textit{being told outright} what they should and should not believe. This engages what Kahneman \cite{kahneman2011thinking} calls ``System 1'' thinking---effortless, intuitive, and sometimes illogical. Provenance, however, takes a different approach: provide users with resources they need to make their own informed decisions on whether to believe what they see. In other words, users still need to \textit{make their own judgements} rather than having one prescribed to them. Kahneman calls this more effortful and rational style of thinking ``System 2.'' Design affordances that allow users to better switch between ``System 1'' and ``System 2'' modes of thought while evaluating media can allow them to effectively leverage provenance information to make credibility decisions.

When designing provenance indicators, it is important to balance using familiar visual cues (in line with some participant suggestions in Section \ref{s:qual}), and using a distinct visual language that may improve conceptual understanding. Case studies on user interfaces used when introducing new (now well-known) web conventions, such as the AdChoices program \cite{adchoices}, as well as past work in functional metaphors for end-to-end encryption \cite{demjaha2018metaphors}, may have valuable lessons going forward.

\section{Limitations and Future Work}
This study was primarily quantitative, so we could not capture nuanced interactions with specific elements of the provenance UI or account for different interpretations of survey questions, which may have led to inconsistent definitions of \textsc{trust} and \textsc{perceived accuracy}. A complementary qualitative study with participants completing tasks on a provenance-enabled prototype and being interviewed about their experiences would provide more insights. \add{Our deceptive media were also modified through composition, so we did not observe participants use other means of editing to reason about credibility. We anticipate generative AI to be an increasingly popular method of media creation and manipulation on social media in the coming years due to the public availability of text-to-image models, such as DALL-E \cite{ramesh2021dalle} and Stable Diffusion \cite{rombach2022diffusion}. We are thus excited about the potential of provenance standards to supplement budding frameworks in this space \cite{pai-framework} and provide much-needed transparency and disclosures to viewers of AI-generated content.} 

\addd{Our study only included one representation of provenance---the C2PA standard. Thus, it also included C2PA-specific metaphors and design patterns (e.g., cryptographically signed chains), which may not be available in other provenance standards, such as Arweave \cite{arweave} and Four Corners \cite{fourcorners}. Our findings were therefore unable to speak for all provenance standards. Future work may include a broader range of provenance standards to make a more general claim about how users may operationalize the \textit{conceptual knowledge} derived from general provenance exposure---rather than UI-specific signals---to evaluate media.} Additionally, as mentioned in Section \ref{s:design-implications}, we had relatively simple provenance details panels in this study, which may not accurately reflect real-world UIs once the system is deployed. Future work, perhaps of interest to the computer vision community, can study more complex provenance details panels and develop techniques to highlight edits with more visual semantic relevance to users. Our study also used a social media prototype resembling Twitter. Variation in participants' past experiences with and general perceptions of the platform have affected our results in ways we did not account for, and we hope future work may expand similar studies to other platforms.

We measured participants' media credibility perceptions through two quantitative variables---trustworthiness and perceived accuracy of claim---but there are many other approaches that could also be used, as outlined in Section \ref{s:rw-measures}. Our content was also limited to 9 pieces of media per feed to keep the study completion time reasonable and participant fatigue low. This constrained the combination of levels within \textsc{media topic}, \textsc{edit status}, \textsc{claim agreement}, and \textsc{provenance state} we could have for each piece of media. While we managed to include at least one piece of media in each of those level combinations, we did not have any composited news media, nor did we have any composited invalid media. Our media also had concrete ``ground truth'' claims from Snopes, but other content may not have an easily identifiable claim, or have a claim that can be easily agreed upon. Future work can explore impacts of provenance on media with uncertain or nonexistent claims. 

Finally, this study assumed unfamiliarity with provenance UIs and simulated users' first exposure to them, but did not investigate longitudinal effects or byproducts with a temporal component, such as discussions in comments sections or incentives for content creators. We see these as rich avenues for future work.

\section{Conclusion}
Today, the barrier for creating and publishing misleading visual content on social media is low. Media editing tools are increasingly capable of manipulating media in sophisticated and convincing ways, posing a nearly impossible task for human fact-checkers or even automated algorithms to tease apart truth and deception. Provenance protocols present a promising solution: every time a piece of media is created or edited, a record is automatically stamped into the media's metadata. The provenance information can then be surfaced by front-end clients for end users to inspect and use to make an informed judgement about the media's credibility. 

Empowering users with provenance is appealing in theory, but our study revealed that it can also be challenging in practice. We found that access to provenance information can indeed help users correct perceptions on deceptive and manipulated media, but may decrease trust and perceived accuracy with media claims even with perfectly authentic content. Moreover, an additional layer of complexity is introduced in provenance-enabled systems where credibility of \textit{provenance information} co-exists with credibility of the \textit{content itself}. Our study showed that users are not yet prepared to distinguish the two concepts, which can lead to misinterpretations of credibility. Despite this, the average participant self-reported an adequate level of understanding of the provenance UIs' functionality.

Our study provides concrete design suggestions for UIs of provenance-enabled social media platforms, as well as problems for researchers beyond the immediate human-computer interaction and social computing community. Technical work on provenance protocol infrastructure and empirical studies on provenance usability can and should be synergetic. Interdisciplinary work is crucial for addressing key usability challenges to pave the way for successful deployment of usable provenance online.


\begin{acks}
We warmly thank all our participants and reviewers. We also extend a special thanks to Dave Kozma for engineering assistance on the prototype, as well as members of the C2PA Technical Working Group, C2PA UX Taskforce, and the Social Futures Lab for discussion and feedback. This work was supported by a research grant from the C2PA.
\end{acks}

\bibliographystyle{ACM-Reference-Format}
\bibliography{refs}


\appendix
\newpage
\section{All Media Used}
\begin{figure}[h]
    \centering
    \includegraphics[width=1\textwidth]{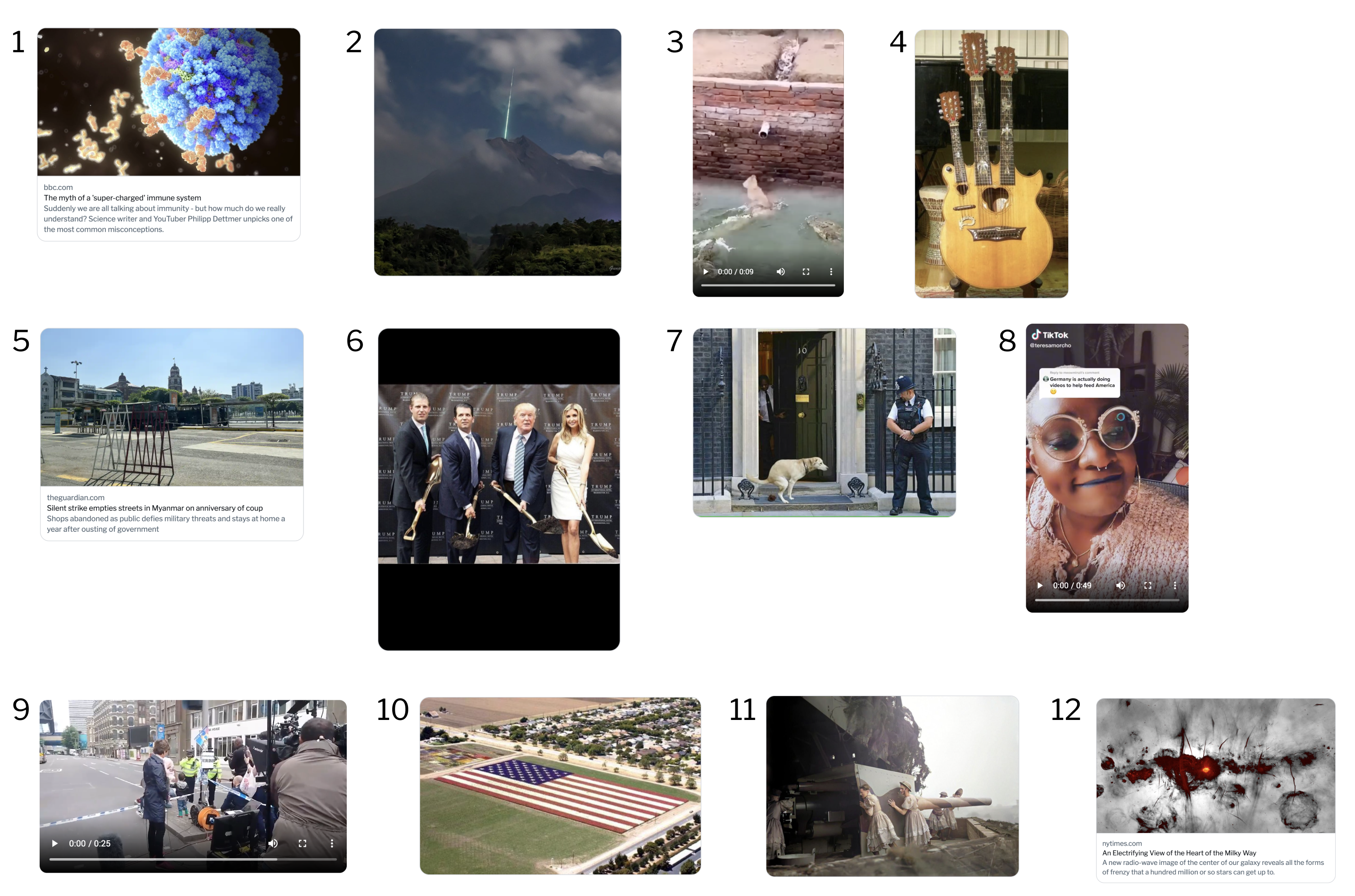}
    \caption{The 12 pieces of media used in the study.}
    \label{fig:all-media}
\end{figure}

\begin{table}
\centering
    \begin{tabular}[h]{p{0.3cm} p{1.1cm} p{0.9cm} p{1.5cm} p{4.2cm} p{1cm} p{1cm} p{1cm}}
    \toprule
    ID & Topic & Edit Status & State & Claim & Claim Agreement & Image/ Video & Region\\
    \midrule 
    1 & News & NC & Incomplete & This is an artistic rendering of cells in the human immune system. & Agree & Image & Both\\
    2 & Lifestyle & NC & Normal & This is a real photo that captures a meteor falling into an active volcano. & Disagree & Image & Both\\
    3 & Lifestyle & C & Normal & This is a real video that shows a dog narrowly escaping an alligator pit. & Disagree & Video & Both\\
    4 & Lifestyle & NC & Normal & This is a real photo of a "triple-neck" guitar owned by Led Zeppelin bassist John Paul Jones. & Agree & Image & Both\\
    5 & News & NC & Invalid & This is a real photo showing some trucks on an empty street in Myanmar. & Agree & Image & Both\\
    6 & National & C & Incomplete & This is a real photo showing Eric Trump holding a shovel backwards. & Disagree & Image & US\\
    7 & National & C & Incomplete & This is a real photo showing a dog defecating on the steps of 10 Downing Street. & Disagree & Image & UK\\
    8 & National & NC & Invalid & Germany, China, and other countries created a "Great Nations Eat" campaign to feed hungry Americans. & Disagree & Video & US\\
    9 & National & NC & Invalid & CNN created "fake news" and staged protests in which Muslim residents purportedly protested a 3 June 2017 attack in London. & Disagree & Video & UK\\
    10 & National & NC & Normal & This is a real photo that shows a 740' x 390' floral flag made of larkspur. & Agree & Image & US\\
    11 & National & NC & Normal & This is a real historical photograph that shows British troops in drag pointing a naval gun at incoming Nazi planes. & Agree & Image & UK\\
    12 & News & NC & Normal & This is a real radio-wave image of the Milky Way's center. & Agree & Image & Both\\
    \bottomrule
    \end{tabular}
    \caption{Table of all media used (IDs in Fig. \ref{fig:all-media}) with their respective media properties. Under the Edit Status column, \textbf{C} = composited and \textbf{NC} = non-composited.}
    \label{t:media-table}
\end{table}
\label{a:all-media}

\newpage
\section{Demographic Breakdown by Region}
\label{a:demo-region}
\begin{table}[h]
\parbox[t]{.45\linewidth}{
\centering
    \begin{tabular}[t]{p{4.5cm} p{1.5cm}}
    \toprule
    Demographics (US) & Percentage ($n=298$)\\
    \midrule 

    \multicolumn{2}{l}{\textbf{Age}} \\
    \quad 18--24 & 16 \\
    \quad 25--44 & 47 \\
    \quad 45--65 & 37 \\

    \multicolumn{2}{l}{\textbf{Highest Education Obtained}} \\
    \quad High school or below & 30 \\
    \quad Associate's degree/vocational training & 14 \\
    \quad Bachelor's degree & 37 \\
    \quad Master's/professional degree & 17 \\
    \quad Ph.D. & 2 \\

    \multicolumn{2}{l}{\textbf{Annual Household Income (USD)}} \\
    \quad < 30,000 & 25 \\
    \quad 30,000–59.999 & 24 \\
    \quad 60,000–89,999 & 19 \\
    \quad 90,000–149,999 & 17 \\
    \quad 150,000+ & 13 \\
    \quad Decline to Respond & 2 \\
    
    \multicolumn{2}{l}{\textbf{Political Affiliation}} \\
    \quad Liberal & 26 \\
    \quad Liberal-leaning & 16 \\
    \quad Moderate & 18 \\
    \quad Conservative-leaning & 23 \\
    \quad Conservative & 16 \\
    
    \multicolumn{2}{l}{\textbf{Race}} \\
    \quad Asian & 4 \\
    \quad Black/African American & 9 \\
    \quad Mixed & 4 \\
    \quad Native American/Pacific Islander & 1 \\
    \quad White & 81 \\
    \quad Other/Decline to Respond & 1 \\

    \multicolumn{2}{l}{\textbf{Gender}} \\
    \quad Woman & 52 \\
    \quad Man & 46 \\
    \quad Non-binary / third gender & 2 \\

    \bottomrule
    \end{tabular}
    \caption{Demographics of US participants.}
    \label{t:demo-us}
}
\quad
\parbox[t]{.45\linewidth}{
\centering
    \begin{tabular}[t]{p{4.5cm} p{1.5cm}}
    \toprule
    Demographics (UK) & Percentage ($n=297$)\\
    \midrule 

    \multicolumn{2}{l}{\textbf{Age}} \\
    \quad 18--24 & 13 \\
    \quad 25--44 & 43 \\
    \quad 45--65 & 44 \\

    \multicolumn{2}{l}{\textbf{Highest Education Obtained}} \\
    \quad High school or below & 41 \\
    \quad Associate's degree/vocational training & 17 \\
    \quad Bachelor's degree & 29 \\
    \quad Master's/professional degree & 10 \\
    \quad Ph.D. & 3 \\

    \multicolumn{2}{l}{\textbf{Annual Household Income (GBP)}} \\
    \quad < 20,000 & 22 \\
    \quad 20,000–39.999 & 32 \\
    \quad 40,000–59,999 & 20 \\
    \quad 60,000–99,999 & 9 \\
    \quad 100,000+ & 2 \\
    \quad Decline to Respond & 15 \\

    \multicolumn{2}{l}{\textbf{Political Affiliation}} \\
    \quad Left wing & 14 \\
    \quad Centre-left & 23 \\
    \quad Centre & 39 \\
    \quad Centre-right & 20 \\
    \quad Right wing & 4 \\
    
    \multicolumn{2}{l}{\textbf{Race}} \\
    \quad Asian & 5 \\
    \quad Black & 3 \\
    \quad Mixed & 2 \\
    \quad White & 89 \\
    \quad Other/Decline to Respond & 1 \\

    \multicolumn{2}{l}{\textbf{Gender}} \\
    \quad Woman & 53 \\
    \quad Man & 46 \\
    \quad Non-binary / third gender & 0.3 \\
    \quad Prefer not to disclose & 0.7 \\

    \bottomrule
    \end{tabular}
    \caption{Demographics of UK participants.}
    \label{t:demo-uk}
}
\end{table}

\newpage
\section{Provenance Indicator Design and Language Variations}
\label{a:design-spread}

\begin{figure}[h]
    \centering
    \includegraphics[width=1\textwidth]{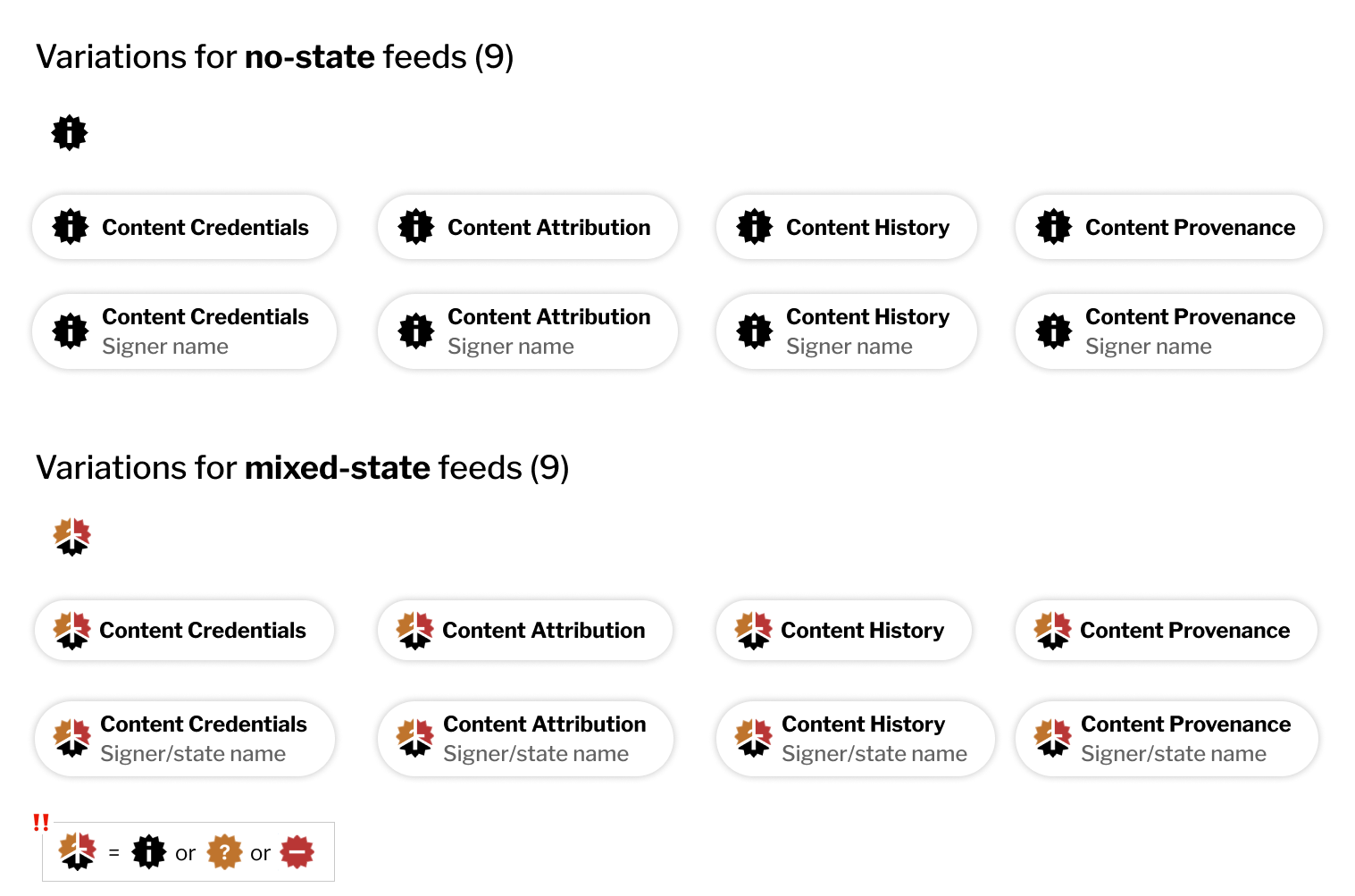}
    \caption{Our 18 variations of provenance indicators across no-state and mixed-state feeds. The split icons in the mixed-state section indicates that the icon can be either the normal, incomplete, or invalid one.}
    \label{fig:design-spread}
\end{figure}

\section{Survey Screenshots}
\label{a:survey}

\begin{figure}[h]
    \centering
    \includegraphics[width=0.9\textwidth]{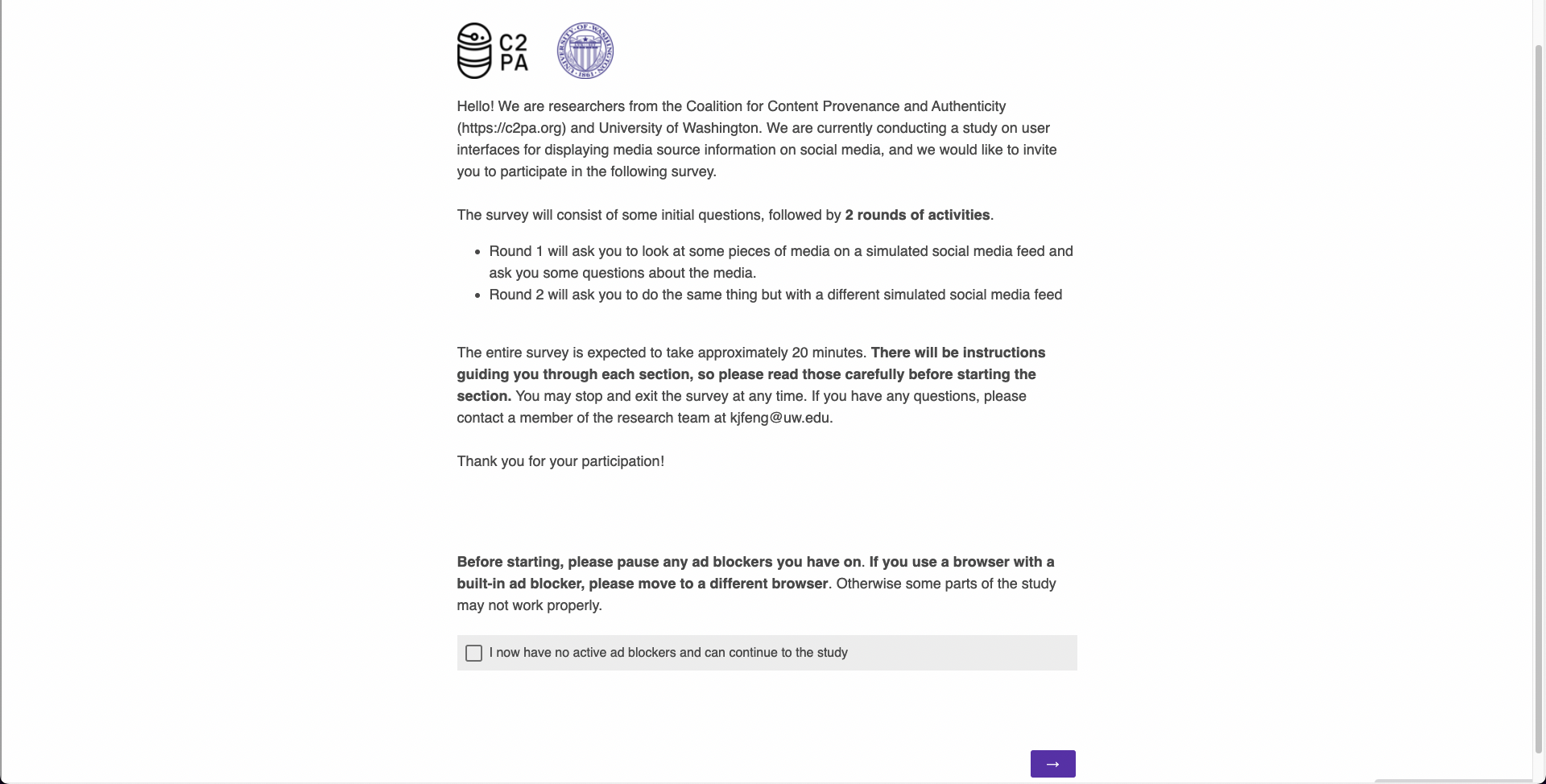}
    \caption{Intro page.}
    \label{fig:s1}
\end{figure}

\begin{figure}[h]
    \centering
    \includegraphics[width=1\textwidth]{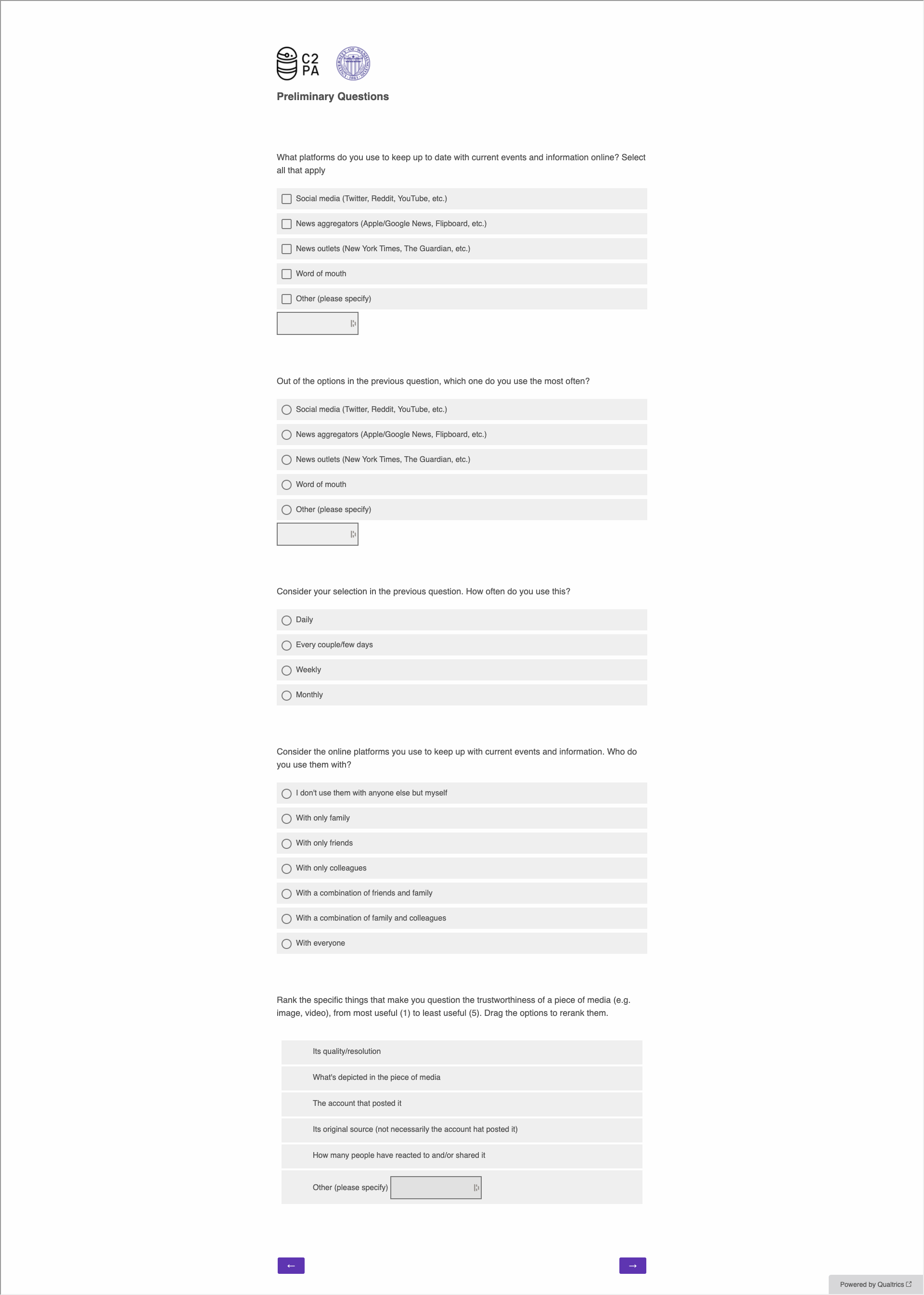}
    \caption{Preliminary questions about social media usage patterns.}
    \label{fig:s2}
    
\end{figure}

\begin{figure}[h]
    \centering
    \includegraphics[width=1\textwidth]{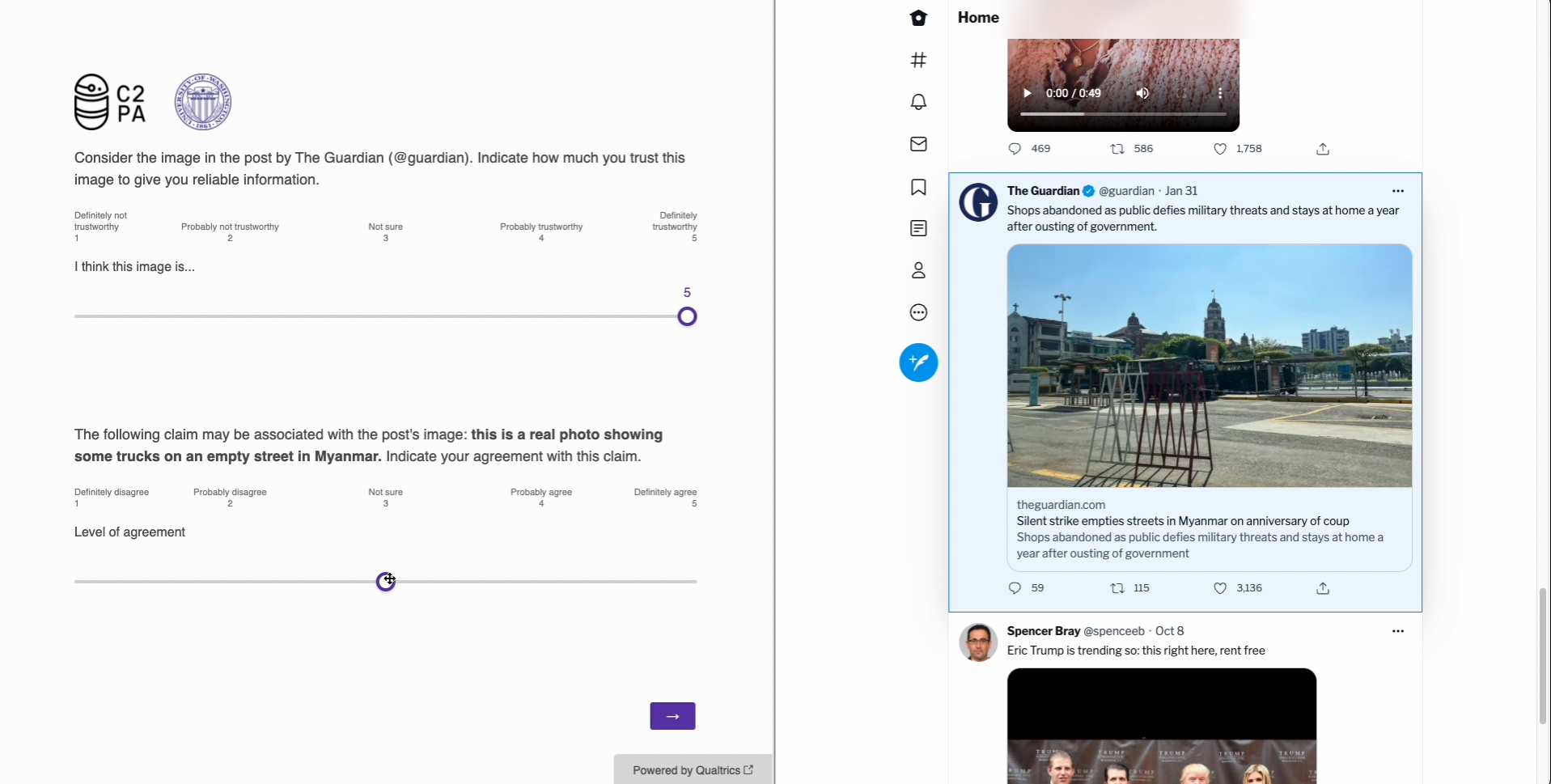}
    \caption{First round of media evaluations.}
    \label{fig:s3}
    
\end{figure}

\begin{figure}[h]
    \centering
    \includegraphics[width=1\textwidth]{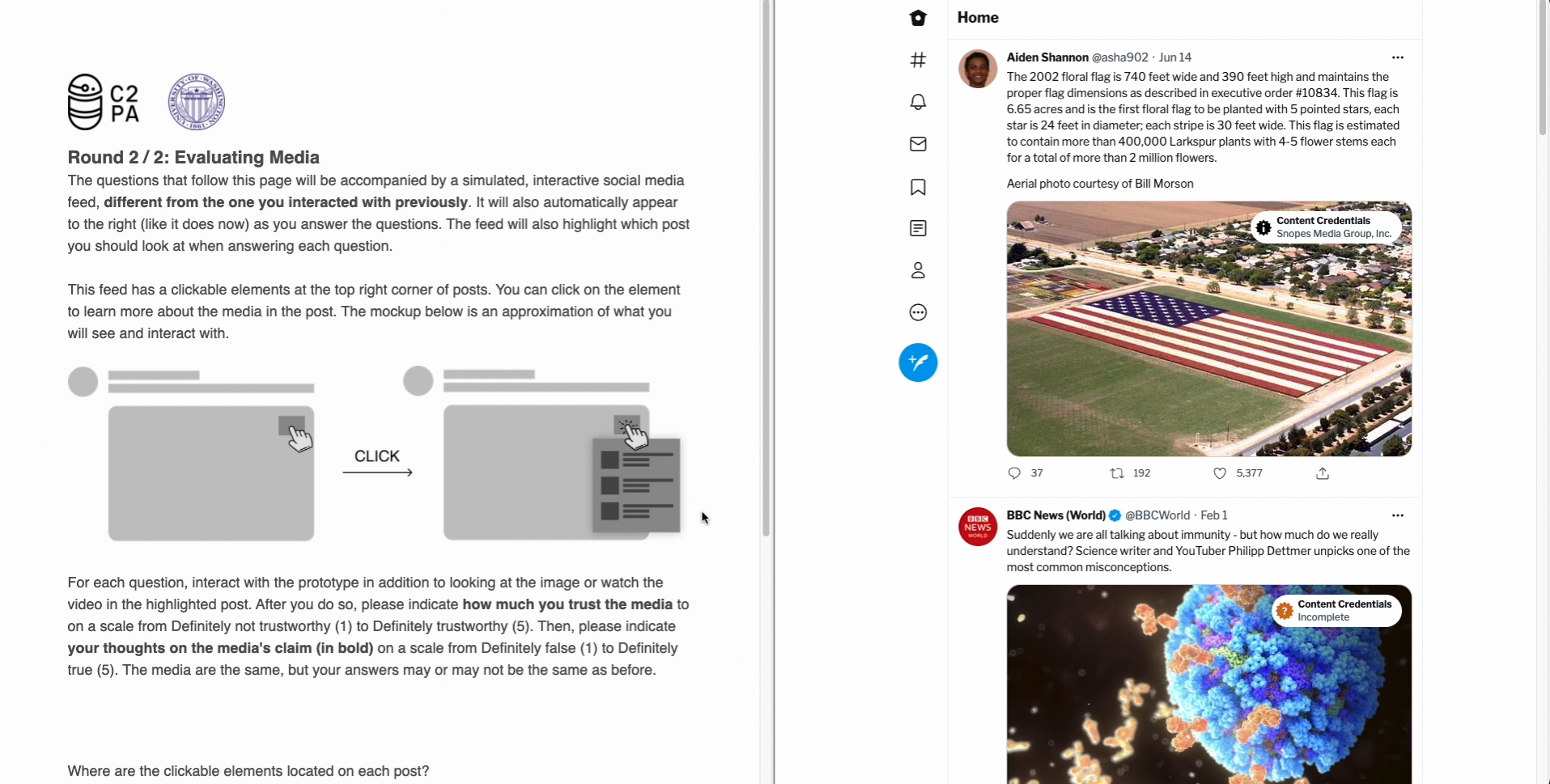}
    \caption{Graphical debrief between first and second rounds of media evaluations.}
    \label{fig:s4}
\end{figure}

\begin{figure}[h]
    \centering
    \includegraphics[width=1\textwidth]{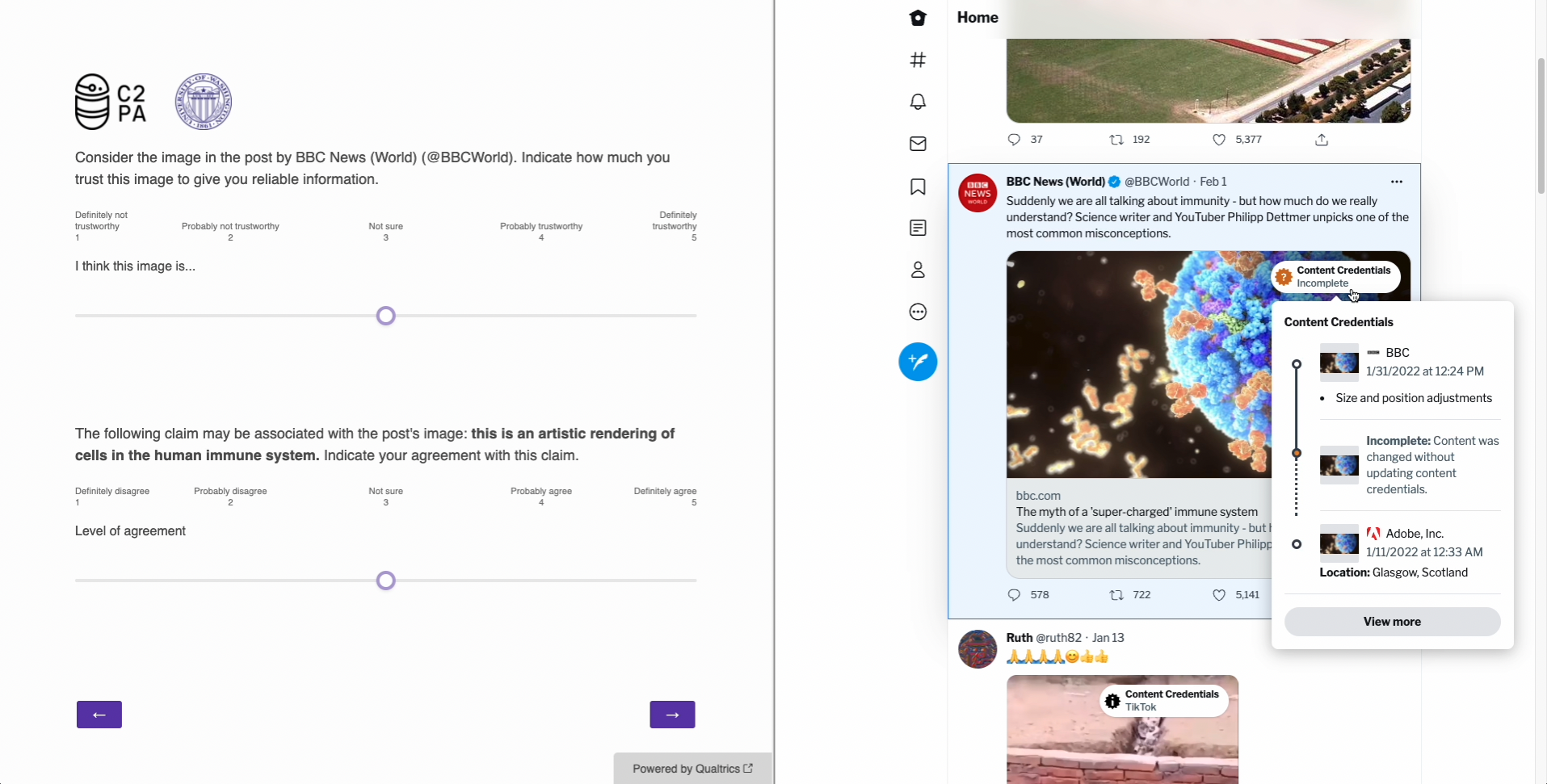}
    \caption{Second round of media evaluations.}
    \label{fig:s5}
\end{figure}

\begin{figure}[h]
    \centering
    \includegraphics[width=1\textwidth]{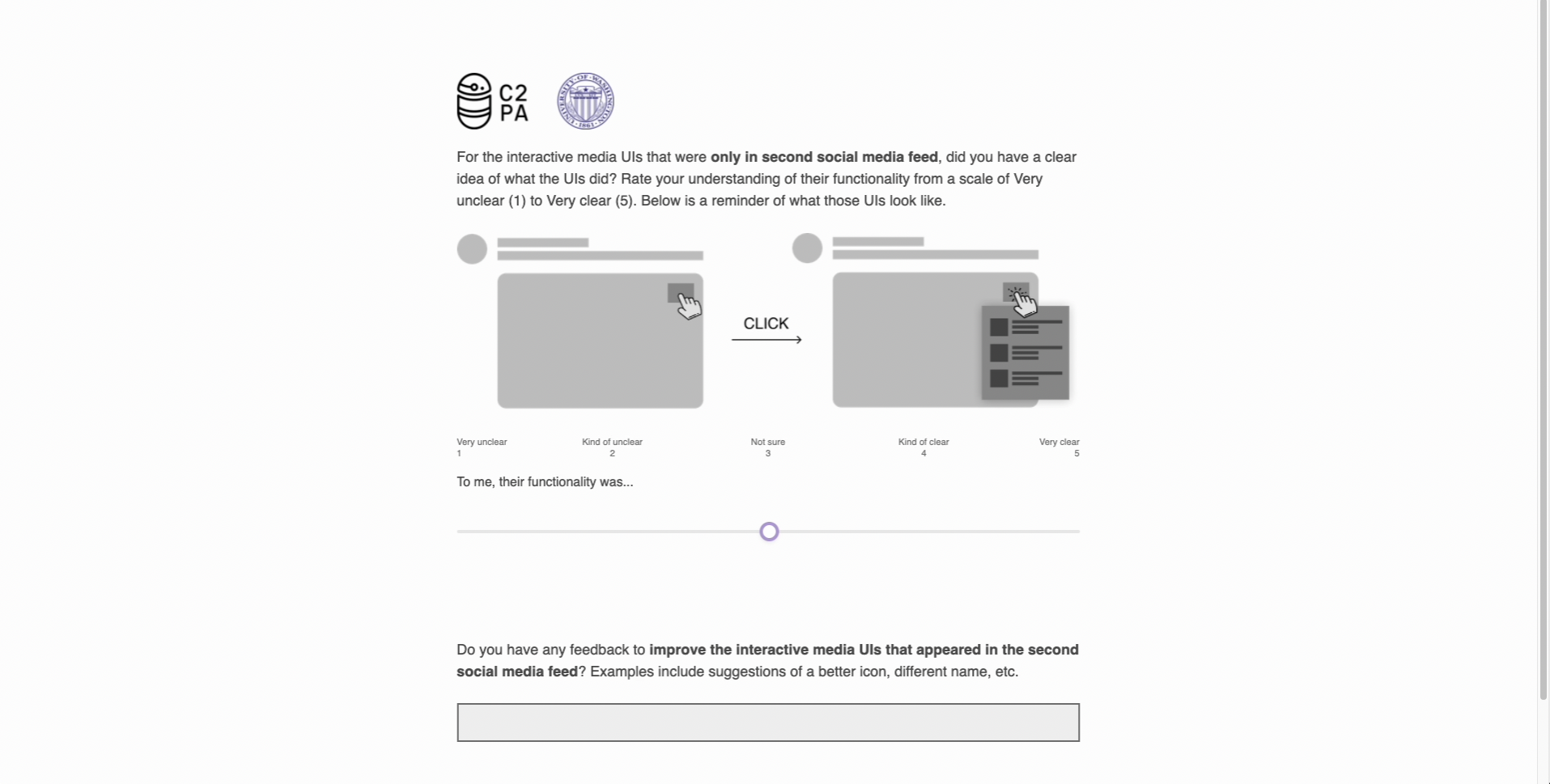}
    \caption{Exit survey part 1: capturing comprehension and feedback.}
    \label{fig:s6}
\end{figure}

\begin{figure}[h]
    \centering
    \includegraphics[width=1\textwidth]{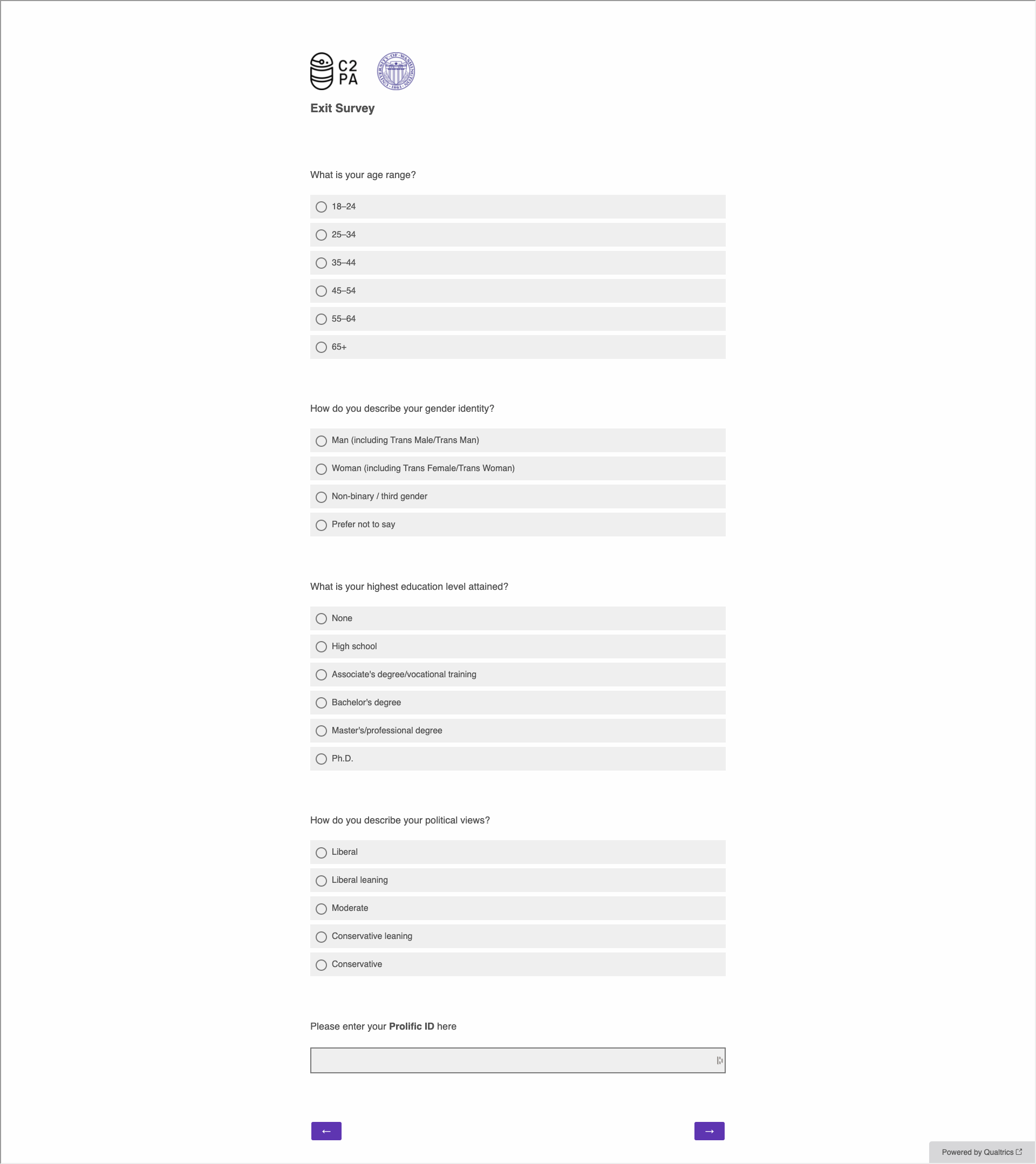}
    \caption{Exit survey part 2: demographics.}
    \label{fig:s7}
\end{figure}

\begin{figure}[h]
    \captionsetup{justification=raggedright, singlelinecheck=false}
    \caption*{Received July 2022; revised January 2023; accepted March 2023.}
\end{figure}

\end{document}